\documentclass{aa}

\usepackage[utf8]{inputenc}

\usepackage[varg]{txfonts}
\usepackage{bm}
\usepackage{mathtools}
\usepackage{amssymb}
\usepackage{graphicx}
\usepackage{epstopdf}
\usepackage{natbib}
\usepackage{color}
\usepackage[colorlinks=true,     linkcolor=blue, citecolor=blue, filecolor=blue, urlcolor=blue]{hyperref} 
\usepackage{float}
\bibpunct{(}{)}{;}{a}{}{,}

\usepackage{multirow,multicol}
\usepackage{booktabs}

\newlength{\verticalcompensationlength}
\newcounter{verticalcompensationrows}

\usepackage{lscape}
\usepackage{rotating}
\usepackage[normalem]{ulem}
\pdfoutput=1
\begin{document}

   \title{How dust particles orbiting white dwarfs can reveal undetected co-orbiting exoplanets}
\subtitle{II. Exoplanets on elliptic orbits}

 \author{Kyriaki I. Antoniadou\inst{\ref{inst1},\ref{inst2}} \and   Dimitri Veras\inst{\ref{inst3},\ref{inst4},\ref{inst5}}
          }

   \institute{Department of Physics, Democritus University of Thrace, 65404, Kavala, Greece
   \\ \email{kantonia@physics.duth.gr}\label{inst1} 
   \and
    Department of Physics, Aristotle University of Thessaloniki, 54124, Thessaloniki, Greece \label{inst2}
   \and
    Centre for Exoplanets and Habitability, University of Warwick, Coventry CV4 7AL, United Kingdom \label{inst3}
   \and
    Centre for Space Domain Awareness, University of Warwick, Coventry CV4 7AL, United Kingdom\label{inst4}
   \and
        Department of Physics, University of Warwick, Coventry CV4 7AL, United Kingdom\label{inst5}
         }

\titlerunning{II. Observational hints about debris}

\authorrunning{K. I. Antoniadou and D. Veras}

\date{Received XXXX; Accepted YYYY}

  \abstract
        {The time evolution of the observed periodicities of transiting rocky debris around white dwarfs (WDs) potentially signposts ongoing dynamical interactions between the debris and co-orbital asteroids or planets with similar orbital periods (1/1 mean-motion resonance). In Paper I, we analysed these WD-secondary-dust particle systems within the framework of the circular restricted three-body problem (CRTBP).}   
     {We aim to extend Paper I to the eccentric secondary orbit case and now analyse the elliptic restricted three-body problem (ERTBP). Our secondary masses include Ceres, Mercury, Earth, and Neptune so that we can help infer the presence of both large asteroids and planets through our analysis. We study dust particles on both planar (2D-ERTBP) and non-coplanar (3D-ERTBP) orbits. }
   {We computed families of 1/1 resonant symmetric and asymmetric periodic orbits in the 2D-ERTBP and 3D-ERTBP. Doing so produces novel results in their own right and diagnostic tools for precise guidance for our next steps. Based on them, we generated detailed phase portraits. Finally, we run suitably seeded $N$-body simulations to compare with the periodic orbits, and to provide observational links with debris close to WDs.}
   {New families of asymmetric periodic orbits, called $E^A_{41}$, $E^A_{42}$, and $E^A_{43}$ in the high-eccentricity 2D-ERTBP regime were discovered. They are generated by the new bifurcation point, $B_T^4$, along the new $A_2$ family, showcased in Paper I. Moreover, new isolated families of asymmetric periodic orbits, namely $E^A_{44}$, $E^A_{45}$, and $E^A_{46}$ were discovered. Families of short- and long- period asymmetric orbits starting from $L_4$, called $E^A_S$ and $E^A_L$, were also computed.} 
   {In both the 2D- and 3D-ERTBP regimes, highly eccentric motion is exhibited for the dust particles in the majority of the families of either symmetric or asymmetric periodic orbits. The portraits reveal stability domains as a function of the dust’s eccentricity, argument of pericentre, mean anomaly and longitude of ascending node with slight semi-major axis deviations from the 1/1 MMR. The $N$-body simulations reveal that over ten years, the orbital period deviations of the dust are higher than in the CRTBP; in some architectures, Mercury-mass planets can regularly generate deviations exceeding $\sim$100 s, which potentially aids observational inferences of hidden asteroids or planets orbiting WDs.}
   \maketitle
\nolinenumbers

\section{Introduction} \label{intro}

\begin{figure*}
$\begin{array}{c}
  \includegraphics[width=0.99\textwidth]{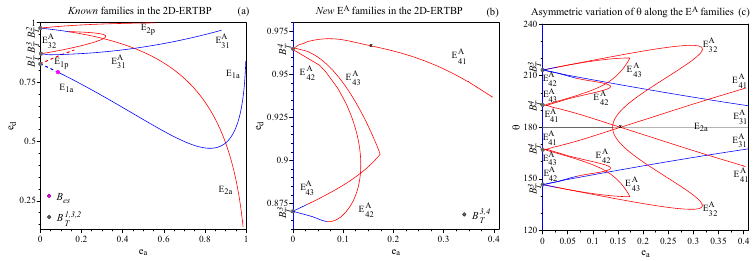}\vspace{-0.3cm}
\end{array}$
\caption{Families of planar symmetric $E$ and asymmetric $E^A$ periodic orbits on the $(e_{\rm a},e_{\rm d})$ plane (\textit{panels a and b}) and on the $(e_{\rm a},\theta)$ plane (\textit{panel c}) in the 2D-ERTBP. Subscript ${\rm d}$ denotes the location of the asteroid at pericentre, while ${\rm a}$ represents the location of the asteroid at apocentre at $t=0$. In panels (b) and (c), the new families, $E^A_{41}$, $E^A_{43}$, and $E^A_{43}$, of asymmetric periodic orbits generated by $B^4_T$ are demonstrated in colour, while the black x-symbol along the $E^A_{41}$ family of asymmetric orbits identifies a crossing orbit on this projection, where $\theta=180^{\circ}$, without it being a bifurcation point of the $E_{2a}$ family of symmetric periodic orbits. The magenta dot, $B_{es}$, shown in panel (a) generates periodic orbits in the 3D-ERTBP (see also Fig. \ref{3DE} and Table \ref{tab1} for details about the full configuration). Stable (unstable) periodic orbits are blue (red). Solely the transition of stability along the $E_{2a}$ family of symmetric periodic orbits is depicted by different tones of grey in panel (c). Vertical instability is only found along the $E_{1a}$ and $E_{1p}$ families in panel (a) and is depicted by dashed lines.}\vspace{-0.3cm}
        \label{2DE}
\end{figure*}

As the number of extreme planetary systems increases, the analytical and numerical tools used for characterisation and prediction need to keep up. While periodic orbits, dynamical stability maps, and the restricted three-body problem (RTBP) have been regularly applied to extra-solar planetary systems around main-sequence stars, white dwarf (WD) planetary systems pose different challenges and are not as well-studied.

In particular, in these systems both the mass parameter, $\mu$, and the compactness of orbits can be extreme. Observed transiting debris clouds are signatures of the breakup of minor planets, giving $\mu \sim 10^{-11}-10^{-9}$ at the upper end. Further, nearly all of these clouds have orbital periods of between just four and 25 hours \citep{vander15,vander20,vander21,farihi22,bhat25,guid25,hermes25,korth26}.

Aside from the debris' existence, it is dynamically active on short timescales. The debris activity around the white dwarf WD~1145+017 \citep{vander15} almost completely ceased about nine years after first being detected. During the first few years of measured activity, there were indications of an extant minor planet gravitationally perturbing fragments that likely originated from that minor planet \citep{rapp16,rap18} and corresponding efforts to study them \citep{vmtg16,gurr17,ver17}. More recently, there was a significant event concerning the white dwarf ZTF~J1944+4557; debris that transited at 4.9704 hours disappeared for seven months before reappearing with the same orbital period.

In most of these systems, the debris is assumed (but not known) to have largely circular orbits. A notable exception is the WD ZTF~J0139+5245 \citep{vander20}, where debris is inferred to orbit with an eccentricity exceeding 0.95. Origin theories for the debris predict the existence of these types of systems \citep[e.g.][]{brou22}, which may currently be in a transitional state towards circularisation.

While existing minor planets, whether partially or fully intact, may play a role in the dynamical evolution of transiting debris systems, major planets may also lurk close to the WDs. Terrestrial and giant planets can be difficult to detect, primarily due to the dimness of many WDs. However, the presence of the ice giant around WD~J0914+1914 at a separation of 0.07~au \citep{gans19} and of the gas giant around WD~1856+534 at a separation of 0.02~au \citep{vanderburg20} reveals that large planets do reach these locations, and would certainly help shape debris evolution in such systems. Terrestrial planets have not yet been discovered orbiting within 0.1~au of a white dwarf, but their detection represents a major goal.

In this respect, the debris evolution may indicate the presence of hidden planets. We analysed this possibility in two previous papers, \cite{ave24} (P3 hereafter) and \cite{ave26a} (P4 hereafter); these studies applied the same techniques as our first two papers, \cite{ave16} (P1 hereafter) and \cite{ave19} (P2 hereafter). All five works, which are summarised in Table A.1 of P4, applied periodic orbits, dynamical stability (DS) maps, and $N$-body simulations to white dwarf planetary systems. P3, P4, and this work (P5) specifically investigated co-orbital debris with asteroids and planets.

Both P1 and P2 revealed potential locations of hidden planets at astronomical-unit-scale distances orbiting white dwarfs whose atmospheres contain observable debris. P3 instead analysed the situation much closer to the white dwarf and revealed potential locations of hidden asteroids at solar-radius-scale distances orbiting white dwarfs that contain debris observed by photometric transits. P3 only analysed one asteroid mass, which was equal to 0.1 Ceres mass. However, much larger objects may be lurking close to white dwarfs. Consequently, both P4 and P5 extended P3 to higher secondary mass cases, with four mass parameters for Neptune-like ($\mu=8.58\times 10^{-5}$), Earth-like ($\mu=5.005\times 10^{-6}$), Mercury-like ($\mu=2.76\times 10^{-7}$), and Ceres-like ($\mu=7.86\times 10^{-10}$) objects orbiting white dwarfs.
However, P4 was limited to investigating a secondary on circular orbit, namely the 2D-CRTBP and 3D-CRTBP. The assumption of circularity is not necessarily realistic in all cases, as demonstrated by, for example, the white dwarf planetary system ZTF~J0139-5245 \citep{vander20}, where the disintegrating secondary appears to be in a highly eccentric orbit. In this work (P5),  we adopted the same mass parameters as in P4 but lifted the major restriction of circularity and analysed a secondary on an elliptic orbit (i.e. the 2D-ERTBP and 3D-ERTBP). 

The paper is organised as follows. In Sect. \ref{POs}, we provide the periodicity conditions for periodic orbits, which can be coplanar (Sect. \ref{2dertbp}) or mutually inclined (Sect. \ref{3dertbp}). In App. \ref{app1}, we provide the bifurcation points that generate periodic orbits in the ERTBP. In App. \ref{app2}, we present the continuation of families of periodic orbits in the Ceres-Neptune regime of the 3D-ERTBP. In Sect. \ref{maps}, we compute DS maps as a function of the most important combinations of orbital elements, and we showcase how the evolution of the angles may trap the dust particles around WDs in App. \ref{app3}. In Sect. \ref{sims}, we perform $N$-body simulations. We discuss our findings in Sect. \ref{dis} before concluding in Sect. \ref{concl}.

\begin{figure}
$\begin{array}{c}
  \includegraphics[width=0.94\columnwidth]{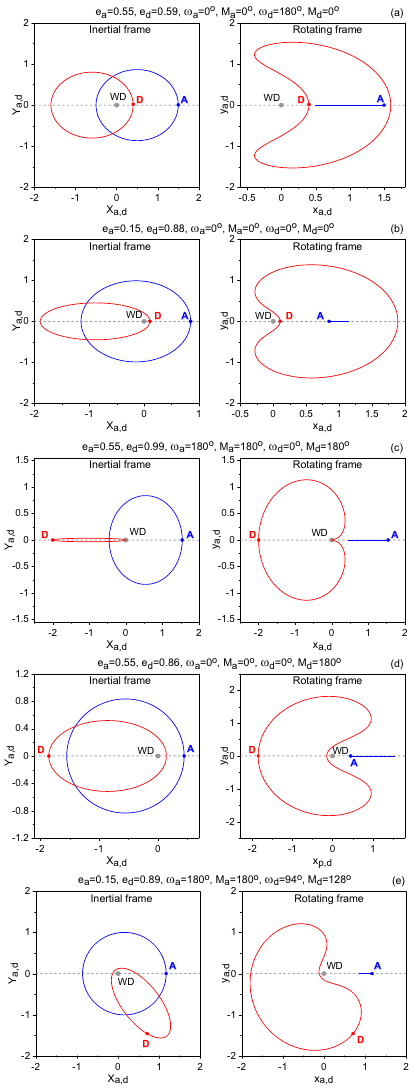}\vspace{-0.325cm}
\end{array}$
\caption{Periodic orbits extracted from the known families, namely $E_{1a}$ (\textit{panel a}), $E_{1p}$ (\textit{panel b}), $E_{2a}$ (\textit{panel c}), $E_{2p}$ (\textit{panel d}), and $E^A_{32}$ (\textit{panel e}), integrated for one period and presented in the inertial and the rotating frames of reference. The initial conditions related to Fig. \ref{2DE}a are reported above each row, while $A$ (blue dot) and $D$ (red dot) denote the asteroid and the dust particle. The WD is represented by a grey dot.\vspace{-0.25cm}}
        \label{sorbits}
\end{figure}

\begin{figure}
$\begin{array}{c}
  \includegraphics[width=0.94\columnwidth]{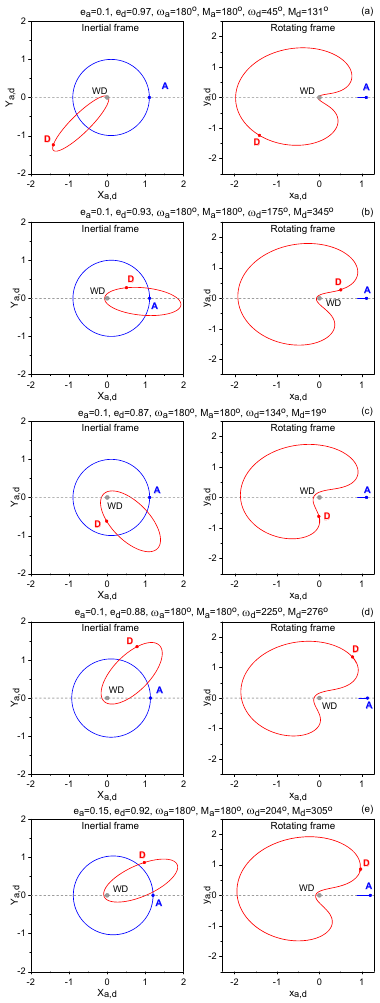}\vspace{-0.325cm}
\end{array}$
\caption{Unstable asymmetric periodic orbits extracted from the new $E^A$ families $E^A_{41}$ (\textit{panel a}),  $E^A_{42}$ (panels b and c), and $E^A_{43}$ (panels d and e). The initial conditions related to Fig. \ref{2DE}b,c are reported above each row. The presentation is as in Fig. \ref{sorbits}.} 
        \label{anorbits}
\end{figure}

\begin{figure*}
$\begin{array}{c}
  \includegraphics[width=0.99\textwidth]{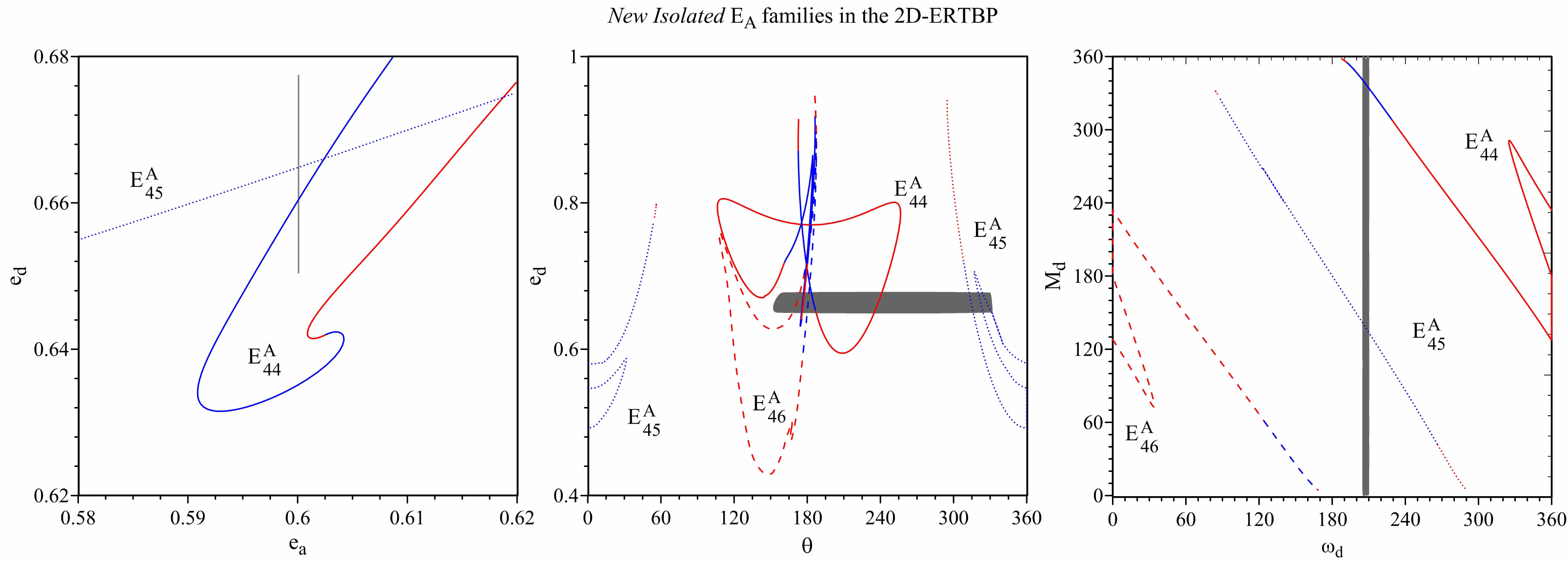}\vspace{-0.25cm}
\end{array}$
\caption{New isolated (not linked with a bifurcation point) families of asymmetric periodic orbits, namely $E^A_{44}$ (solid line) and $E^A_{45}$ (dotted line), while $E^A_{46}$ (dashed line) is the mirror image of $E^A_{44}$. Blue corresponds to stability, while red corresponds to instability. The grey curve corresponds to the evolution of an Earth-like planet  with the initial condition $(e_d,\omega_d)\approx (0.65,152^{\circ})$ or $(0.65,208^{\circ})$ (see also the associated island in the third column of Fig. \ref{2DE_TB}). In the left panel, we magnify the dynamical neighbourhood on the eccentricity plane while showcasing the variation of the angles in the rest of the panels along the full extent of the families. The right panel justifies the fact that $E^A_{46}$ is the mirror image of $E^A_{44}$ (or vice versa).}\vspace{-0.2cm}
        \label{earth}
\end{figure*}

\begin{figure*}
$\begin{array}{c}
  \includegraphics[width=0.99\textwidth]{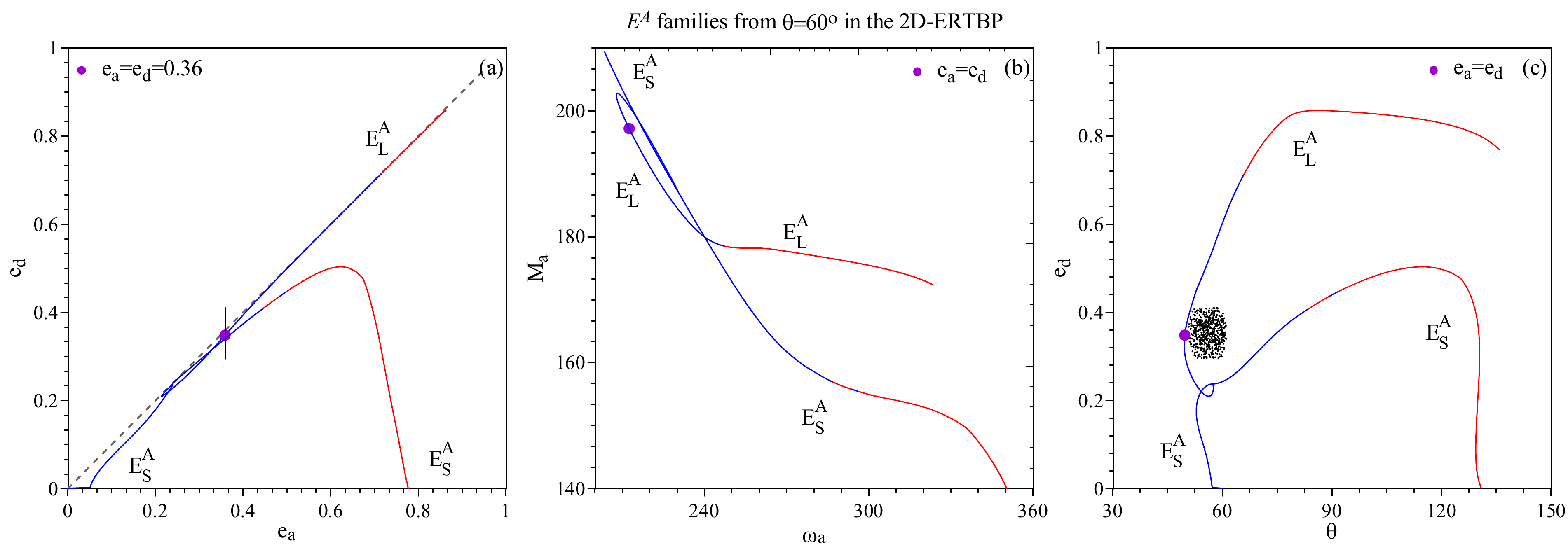}\vspace{-0.25cm}
\end{array}$
\caption{Families of asymmetric periodic orbits in the 2D-ERTBP starting from $\theta\approx 60^{\circ}$ presented on the  $(e_{\rm a},e_{\rm d}) $ (\textit{panel a}), $(\omega_{\rm a},M_{\rm a})$ (\textit{panel b}), and  $(\theta,e_{\rm d})$ (\textit{panel c}) planes. $E^A_S$ corresponds to the family of short-period asymmetric periodic orbits, while $E^A_L$ corresponds to the long-period ones. Blue stands for stability and red for instability. The dashed grey line represents the location of $e_{\rm a}=e_{\rm d}$. The black curve in panel (a) is the evolution of an Earth-like planet centred at a vertically stable periodic orbit with initial conditions of $e_{\rm a}=e_{\rm d}=0.36$ (purple dot) and is hosted in the vertically stable island (invariant torus) shown in the second row of Fig. \ref{3DE_Di45}. The  black evolution shown in panel (c) is bounded within the island of stability of that DS map.}\vspace{-0.25cm}
        \label{aL4}
\end{figure*}

\section{Definition of periodic orbits}\label{POs}

Here we consider a primary body denoted WD; we use the subscript ${\rm a}$ to denote the secondary body (exoplanet or asteroid)   and the subscript ${\rm d}$ to denote the third massless body (dust particle) moving in a suitable rotating frame or reference, $Oxyz$ (see \citealt{sze} for more details about its exact definition). For consistency, this nomenclature is equivalent to that used in P3 and P4. In all cases, the system obeys the fundamental symmetry $\Sigma: (t,x,y) \rightarrow (-t,x,-y)$ \citep{hen97}.

A periodic orbit is symmetric with respect to the $x$-axis when it crosses the $Ox$-axis perpendicularly twice. Therefore, the symmetric periodic orbits with respect to the $x$-axis in the 2D-ERTBP fulfil the periodicity conditions
\begin{equation}\begin{array}{lll}
        x_{\rm a}(T)=x_{\rm a}(0),& x_{\rm d}(T)=x_{\rm d}(0), & \dot{y}_{\rm d}(T)=\dot{y}_{\rm d}(0), 
\end{array}\label{one}\end{equation}
where $T$ is the orbit's period satisfying $t=kT$, with $k\geq 1$ being an integer. At the symmetric periodic orbits the longitude of pericentre, $\varpi$,  equals $0^{\circ}$ or $180^{\circ}$, and the body may be initially located at the pericentre or the apocentre through the value of the mean anomaly, $M$. Therefore, we may have aligned ($\Delta \varpi=0^{\circ}$) or anti-aligned ($\Delta \varpi=180^{\circ}$) bodies. The resonant angle, defined here as $\theta=\lambda_{\rm a}-\lambda_{\rm d}$ \citep{murray}, may librate about $0^{\circ}$ or $180^{\circ}$ around stable symmetric periodic orbits and other values when the periodic orbits are asymmetric.

The apsidal difference, $\Delta \varpi$, is a measure of geometric asymmetry, while the difference between the mean anomalies, $\Delta M$, is a measure of dynamic asymmetry. The asymmetric periodic orbits in the 2D-ERTBP may be defined in the following two ways: 
\begin{itemize}
\item{If the orbit of the dust starts on the plane $(x,y),$ and provided that $\dot{x}_{\rm a}(T)=\dot{x}_{\rm a}(0)=0,$ the initial conditions would be
\begin{equation}\begin{array}{lll}
        x_{\rm a}(T)=x_{\rm a}(0),& x_{\rm d}(T)=x_{\rm d}(0),&y_{\rm d}(T)=y_{\rm d}(0),\\
      &  \dot{x}_{\rm d}(T)=\dot{x}_{\rm d}(0),&\dot{y}_{\rm d}(T)=\dot{y}_{\rm d}(0)\neq0.
\end{array}\label{two}\end{equation}
}
\item{If the orbit of the secondary does not start perpendicularly from the $Ox$-axis, the asymmetric periodic orbits (provided that the dust particle starts on the $Ox$-axis,  $y_{\rm d}(T)=y_{\rm d}(0)=0)$ are represented in the ERTBP by the following conditions:
\begin{equation}\begin{array}{lll}
        x_{\rm a}(T)=x_{\rm a}(0),&x_{\rm d}(T)=x_{\rm d}(0), &\\
        \dot {x}_{\rm a}(T)=\dot {x}_{\rm a}(0)\neq0, &        \dot{x}_{\rm d}(T)=\dot{x}_{\rm d}(0), &\dot{y}_{\rm d}(T)=\dot{y}_{\rm d}(0). \vspace{-0.1cm}\\
\end{array}\label{three}\end{equation}
}
\end{itemize}
Equations \ref{one}--\ref{three} reduce to the periodicity conditions of the 2D-CRTBP studied in P4 when $x_{\rm a}=1-\mu$ (constant), where $\mu=\frac{m_{\rm a}}{m_{\rm a}+m_{\rm WD}}$ is the mass parameter of the problem. Each asymmetric periodic orbit has a mirror image at $360^{\circ}-\Delta\varpi$ and $360^{\circ}-\Delta M$ as a result of the fundamental symmetry.
    
Considering the ERTBP, where the semi-major axis, $a_{\rm a}$, of the secondary is fixed by the normalisation of units at $a_{\rm a}=1$, the variable $x_{\rm a}$ is related to the eccentricity of the secondary and its location at pericentre or apocentre. Examples of known symmetric and asymmetric periodic orbits are presented in Fig. \ref{sorbits}, while examples of asymmetric periodic orbits extracted from the new families fulfilling conditions in Eq. \refeq{two} are shown in Fig. \ref{anorbits}.

The symmetric periodic orbits in the 3D-ERTBP may now be of two\footnote{According to the axial symmetry in plane, two more symmetries may be defined with respect to the $y$-axis and the $yz$-plane \citep[see][]{thesis}, which may increase depending on the value of $\mu$ \citep[][]{mich79c}.}  symmetries:\vspace{-0.1cm}
\begin{itemize}
\item{If the periodic orbit is symmetric with respect to the $xz$-plane, the initial conditions should be \vspace{-0.1cm}
\begin{equation}
\begin{array}{llll}
x_{\rm a}(T)=x_{\rm a}(0),& x_{\rm d}(T)=x_{\rm d}(0),& y_{\rm d}(0)=0,&  z_{\rm d}(T)=z_{\rm d}(0),  \\
\dot x_{\rm a}(0)=0,&\dot x_{\rm d}(0)=0,& \dot y_{\rm d}(T)=\dot y_{\rm d}(0), &\dot z_{\rm d}(0)=0.
\label{xzsym}
\end{array}
\end{equation} }
\item{If the orbit is symmetric with respect to the $x$-axis, the periodicity conditions are\vspace{-0.1cm}
\begin{equation}
\begin{array}{llll}
x_{\rm a}(T)=x_{\rm a}(0), &x_{\rm d}(T)=x_{\rm d}(0), & y_{\rm d}(0)=0, &z_{\rm d}(0)=0,\\
\dot x_{\rm a}(0)=0, & \dot x_{\rm d}(0)=0, & \dot y_{\rm d}(T)=\dot y_{\rm d}(0), & \dot z_{\rm d}(T)=\dot z_{\rm d}(0).
\label{xsym}
\end{array}
\end{equation}}
\end{itemize}

\begin{figure*}
$\begin{array}{c}
\includegraphics[width=0.99\textwidth]{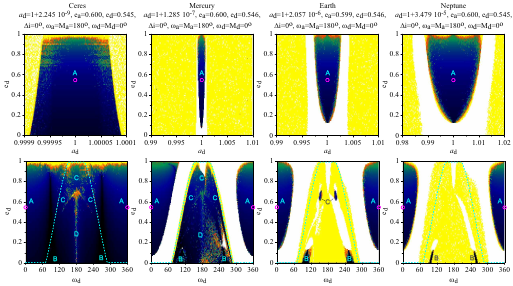}\vspace{-0.1cm}\\ \includegraphics[width=0.18\textwidth]{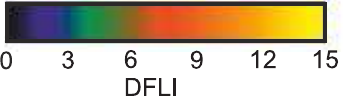}
\end{array}$\vspace{-0.2cm}
\caption{Pairs of DS maps for a secondary body with $e_{\rm a}=0.6$ guided by a stable periodic orbit (magenta circle) of the $E_{Ia}$ family. The full set of the chosen orbital elements are provided at the top of the panels. Dashed cyan curves mark close encounters.  Symbol $\rm A$ stands for the libration of both the resonant angle $\theta$ and the apsidal difference $\Delta\varpi$ around symmetric periodic orbits. Symbol $\rm B$ stands for the libration of $\theta$ and the rotation of $\Delta\varpi$. Symbol $\rm C$ designates areas linked with planar asymmetric periodic orbits, where all angles librate. Symbol $\rm D$ delineates regions where all angles rotate.}
\label{2DE_TB}
\end{figure*}

In this study, we only encountered the $xz$-symmetry along the 3D families of the 3D-ERTBP in Fig. \ref{3DE}, where the values of the ascending node, $\Omega$, of the inclined body equal to $90^{\circ}$ or $270^{\circ}$. Equations \ref{xzsym} and \ref{xsym} reduce to the periodicity conditions of the 3D-CRTBP studied in P4 when $x_{\rm a}=1-\mu$.

\subsection{Secondary on elliptic orbit (ERTBP)}\label{ertbp}
We studied periodic orbits of the first species and the second kind according to \citet{hen97}, thus allowing the eccentricity of the secondary to increase $0<e_{\rm a}\leq 1$, while placing it either at the pericentre or apocentre (subscripts $p$ or $a$ in their names, $E$, of families of symmetric periodic orbits). When there is a change in the geometry of the orbit, we have families of asymmetric periodic orbits (distinguished via a superscript $A$, e.g. $E^A_{31}$). We explored both types of orbits, i.e. coplanar (Sect. \ref{2dertbp}) or mutually inclined (Sect. \ref{3dertbp}).

\subsubsection{Secondary and dust on coplanar orbits (2D-ERTBP)}\label{2dertbp}
In Fig. \ref{2DE}, the families of planar symmetric and asymmetric periodic orbits in the 2D-ERTBP are presented. All families shown in Fig. \ref{2DE}a have been computed by \citet{Pousse17}. Here, as in \citet{vta18}, we added the vertical stability index \citep{hen} and continued the $E_{32}^A$ family up to its termination, which is an orbit of the $E_{2a}$ family, where the linear stability changes at $(e_{\rm a},e_{\rm d})=(0.1393,0.9629)$. In Table \ref{tab1}, the bifurcation points arising along all families are presented.

More precisely, the bifurcation point $B_T^1$ was found within the segment of vertically unstable periodic orbits in Fig. 1a shown in P4. Therefore, both branches, namely $E_{1a}$ and $E_{1p}$, start with vertically unstable orbits (dashed lines). $E_{1a}$ consists of horizontally (linearly) stable orbits, and is the only one where the dust particle can reach moderate eccentricity values $\approx 0.5$, which are appropriate for the $N$-body simulations carried out. More particularly, in the simulations we had to be realistic. Having set $a_{\rm d}=0.012$ au (see Sect. \ref{sims}), we imposed $e_{\rm d}\leq0.6$ in order to avoid the WD’s Roche radius and collisions, which led to the application of the same upper bound for the eccentricity of the  asteroid. More precisely, we needed both $e_{\rm a}$ and $e_{\rm d}$ to be less than 0.6. In this respect, we looked for stable periodic orbits along the families with $max(e_{\rm d})=0.6$ in tandem with $max(e_{\rm a})=0.6$. The other branch, $E_{1p}$, is wholly unstable. The bifurcation point $B_T^2$ can be seen in Fig. 1b in P4 and generates the family $E_{2a}$ --- which is initially stable up to $e_{\rm a}=0.15$ --- and the family $E_{1p}$ of unstable periodic orbits. $B_T^3$ (demonstrated in Fig. 1c in P4) generates the $E_{31}^A$ family of stable asymmetric periodic orbits and the $E_{32}^A$ family of unstable asymmetric ones. We note that all of these families are vertically stable (solid lines). 

In Figs. \ref{2DE}b and c, we demonstrate that the new bifurcation point $B_T^4$ (found in P4) generates two new families of unstable asymmetric periodic orbits, namely $E^A_{41}$, and $E^A_{42}$ which represents a closed family (closes upon itself). More precisely, as $e_{\rm d}$ decreases, $E^A_{42}$ is reflected on $B_T^3$ while accumulating stable asymmetric periodic orbits for $e_{\rm a}<0.067$. Then, as $e_{\rm d}$ increases (we call that segment $E^A_{43}$) it terminates at $B_T^4$. $E^A_{42}$ is not the mirror image of $E^A_{43}$. According to \citet{hen97}, $B_T^3$ is called a reflection point for these two segments ($E^A_{42}$  and  $E^A_{43}$). All asymmetric periodic orbits are vertically stable, but too highly eccentric $(e_{\rm d}>0.6)$ to be suitable for our application to WDs. 

In Fig. \ref{sorbits}, we present some examples of known symmetric and asymmetric periodic orbits presented in Fig. \ref{2DE}a in the inertial and the rotating frames of reference. In Fig. \ref{anorbits}, we showcase some orbits of the new families of asymmetric periodic orbits, $E^A_{41}$ (panel a), $E^A_{42}$ (panels b and  c), and $E^A_{43}$ (panels d and e) in order to demonstrate the reflection at $B_T^3$.

In Fig. \ref{earth}, we present two new isolated (not linked with a bifurcation point) families of asymmetric periodic orbits, namely $E^A_{44}$ and $E^A_{45}$, while $E^A_{46}$ is the mirror image of $E^A_{44}$. These families have been computed by the region of stability revealed for Earth-like planets (third column, bottom panel) in Fig. \ref{2DE_TB} at $(e_d,\omega_d)\approx (0.65,152^{\circ})$ or $(0.65,208^{\circ})$ (see Sect. \ref{maps} for more details). With regard to the computation of the isolated families, we followed a meticulous method. We first validated that $\theta$ librated and confirmed that it exhibited asymmetric libration within those islands. Then, we searched the phase space within that small stability domain for initial conditions that fulfilled the conditions mentioned in Eqs. \ref{two} or  \ref{three}. When we spotted a periodic orbit, we followed the continuation method to form a characteristic curve (family). The concept beyond this method is that in domains of phase space surrounded by stable periodic orbits and thus, populated by invariant tori, the motion is regular and quasi-periodic according to KAM theory \citep{arnold}; \citep[see also e.g.][]{berry,cont}.

In Fig. \ref{aL4}, we demonstrate two families of asymmetric periodic orbits in the 2D-ERTBP starting from $\theta\approx 60^{\circ}$ (see panel c), namely a short-period one called $E^A_S$ and a long-period one called $E^A_L$, which follows a closed loop (shown clearly in panel b). The periodic orbits along $E^A_L$ manifest an evolution in the rotating reference frame similar to the one presented by \citet{Deprit1969}. The dashed grey line added in panel a represents the location of $e_{\rm a}=e_{\rm d}$. The motivation for this computation was the appearance of distinct islands of stability  in the second row of Fig. \ref{3DE_Di45} at $(e_d,\omega_d)\approx (0.36,212^{\circ})$ or $(0.36,328^{\circ})$, namely around $\theta\approx 60^{\circ}$ or $\approx 300^{\circ}$ for Earth-like and Neptune-like secondaries. The black curve in panel a is an example of an evolution centred around a stable periodic orbit of an Earth-like planet with the initial conditions $e_{\rm a}=e_{\rm d}=0.36$ (purple dot) extracted from that DS map (see also Sect. \ref{maps}). This evolution takes place within an invariant torus around the stable periodic orbit, and its centred shape can only be observed when this quasi-periodic evolution is projected on the eccentricity plane. The evolution for the same intial conditions is shown in panel c in relation to the angle $\theta$ and is bounded within the distinct island of stability shown in the DS map.

\subsubsection{Secondary and dust on an inclined orbit (3D-ERTBP)}\label{3dertbp}
In Fig. \ref{3DE}, two families of spatial $xz$-symmetric periodic orbits in the 3D-ERTBP are presented. These families were shown by \citet{va18} for $\mu=10^{-3}$, but here we extended the study to orders of planetary masses $\in[10^{-10}-10^{-5}]$. 

The $F_a$ group of families possesses a secondary located at apocentre, consists of stable orbits and starts from (or ends at) the bifurcation points $B_{es}$/$B_T^5$. It joins the family $F^I$ (also shown in Fig. 4a in P4) to form a bridge of stable, mutually inclined periodic orbits. The other group of families, $F_p$, starts from the bifurcation point $B_T^5$, locates the secondary at pericentre, consists of unstable orbits, and terminates at the moment of collision with the WD at approximately $i_{\rm}=60^{\circ}$.

\begin{figure*}
$\begin{array}{c}
\includegraphics[width=0.97\textwidth]{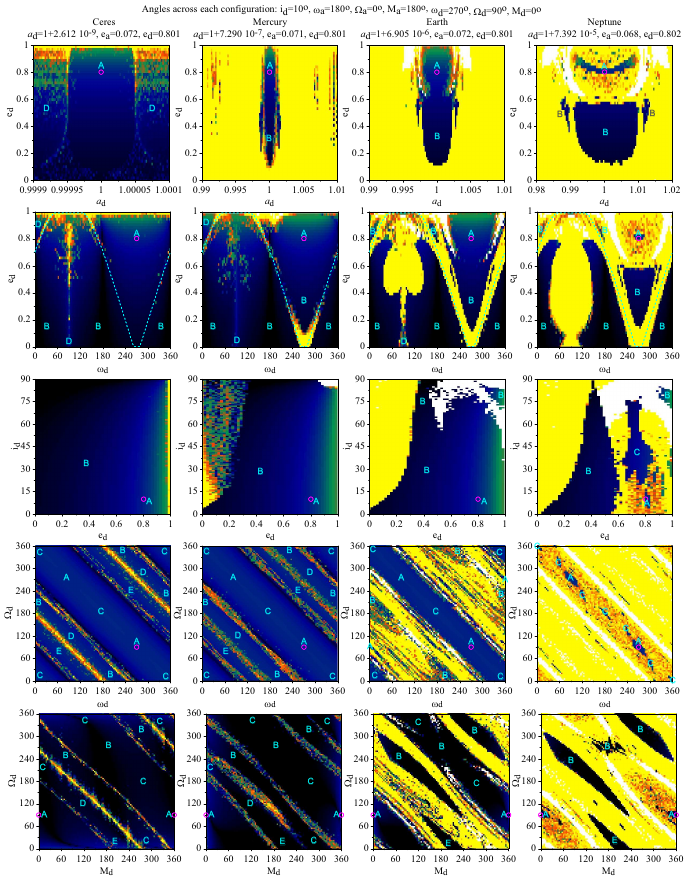}\vspace{-0.25cm}\hspace{0.5cm}\\\includegraphics[width=0.17\textwidth]{bar.pdf}
\end{array}$\vspace{-0.35cm}
\caption{DS maps guided by a 3D stable periodic orbit (magenta circle) of the $F_{a}$ family with $\Delta i=10^{\circ}$, $\Delta\varpi=180^{\circ}$ and $\Delta\Omega=90^{\circ}$, for Ceres, Mercury, Earth, and Neptune in the first to the fourth columns, respectively. The precise initial conditions for these computations are noted in the title of the figure and above each column. Dashed cyan curves denote close encounters. Lettered areas characterise the evolution; namely, in $\rm A$ all angles librate (e-resonance and i-resonance), in $\rm B$ only $\theta$ librates (no MMR), in $\rm C$ only an e-resonance about planar asymmetric orbits exists, in $\rm D$ only the apsidal difference oscillates, and in $\rm E$ all angles rotate.}
\label{3DE_Di10}
\end{figure*}

\begin{figure*}
$\begin{array}{c}
\includegraphics[width=0.99\textwidth]{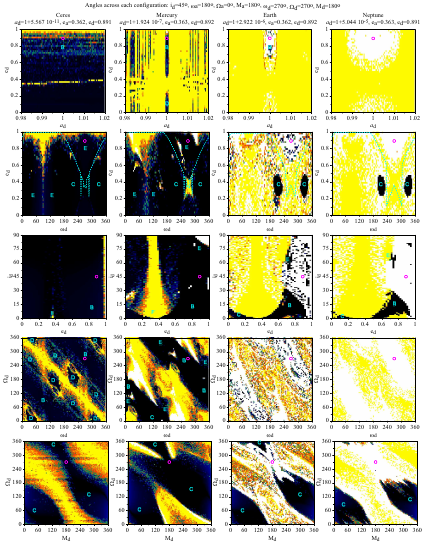}\vspace{-0.25cm}\hspace{0.5cm}\\\includegraphics[width=0.18\textwidth]{bar.pdf}
\end{array}$\vspace{-0.2cm}
\caption{Presented as in Fig. \ref{3DE_Di10} but guided by a 3D unstable periodic orbit (magenta circle) of the $F_{p}$ family with $\Delta i=45^{\circ}$, $\Delta\varpi=0^{\circ}$, and $\Delta\Omega=270^{\circ}$. Symbol $\rm E$ now showcases areas where only the nodal difference oscillates, while symbol $\rm F$ marks an area where all angles rotate.}
\label{3DE_Di45}
\end{figure*}

\subsection{Structure of families in the ERTBP when $\mu\in [10^{-10},10^{-5}]$}
The mass parameter ($\mu$) considered here belongs in the regime $[10^{-10},10^{-5}]$, i.e.  for a Ceres-like up to a Neptune-like secondary, and the differences on the $(e_{\rm a}, e_{\rm d})$ plane are not obvious since this orbital element changes at the fourth significant digit (see the note in Table A.1 and in P4). \citet{va18} presented a list of the known (at that time) bifurcation points, where a difference in the eccentricities was spotted when $\mu$  varied from $10^{-6}$ to $10^{-3}$, i.e. from Earth to Jupiter. In Table 1 of that work, $B_{sce}$ (which we call $B^5_T$ here) changes when $\mu> 10^{-3}$, $B_{es}$ when $\mu> 10^{-4}$ (only for the dust particle and not for the secondary), and $B_{ce}$ (called $B^1_T$ here) when $\mu>10^{-4}$. Even these changes concerned the third significant digit, which was not possible to depict; in particular, the families on the eccentricity plane coincided. 

Therefore, in Figs. \ref{2DE} and \ref{earth}-\ref{aL4}, the structure of the families of periodic orbits represents the expected stationary points of the ERTBP for the mass parameter of the secondary within the  $[10^{-10},10^{-5}]$ range and the third body (dust particle) considered as a massless body.
This is totally different to the General TBP (not studied here), where families of periodic orbits are computed for various values of the mass ratio, $\rho$ (e.g. the mass of the outer planet over that of the inner planet). In that case, we observe a significant difference between the families on the eccentricity plane (see e.g. Fig. 9 and Tables 2 and 3 of \citealt{va18}). 
In the ERTBP, the mass of the dust particle ($m_d$) is equal to zero. Therefore, as the mass of the asteroid or the exoplanet ($m_a$) tends towards zero, $\mu$ will tend towards zero as well. More precisely,  there is no mass ratio ($\rho$) between the two bodies, as we basically approach the unperturbed problem $\mu\rightarrow0$. Hence, the families basically coincide with one another as $\rho=0$ (or infinity) depending on which body is chosen to be the outer. 

The main difference between the families with $\mu$ in the regime $[10^{-10},10^{-5}]$ is the semi-major axis, $a_{\rm d}$, since $a_{\rm a}=1$, which is well noted in the initial conditions written in the top of the DS maps Figs. \ref{2DE_TB}-\ref{3DE_Di45} (and used to suitably seed the $N$-body simulations). In the appendix we demonstrate the families chosen for the simulations. More precisely, in Fig. \ref{E1a_cla}, we showcase the $E_{1a}$ family on the $(a_{\rm d}, e_{\rm d})$ plane, while in Fig. \ref{3deexp} we show the families $F_p$ and $F_a$  and in the $(a_{\rm d}, e_{\rm d}, i_{\rm d})$ space, in order to visualise the minute changes encountered (of the order of $\mu$). 

\subsection{Novelty with respect to the existing literature on families of periodic orbtis in the co-orbital dynamics}
With regard to the 2D dynamics, we computed the linear stability and the vertical stability along the families of periodic orbits and here showcase the vertical critical orbits that generate families in the 3D-ERTBP. The families shown in Fig. \ref{2DE}a, namely $E_{1a}$, $E_{1p}$, $E_{2a}$, $E_{1p}$,  $E_{31}^A$, and $E_{32}^A$ are called $G^{e'}_{QS,1}$, $G^{e'}_{QS,2}$, $G^{e'}_{L_3,1}$, $G^{e'}_{L_3,2}$,  $G^{e'}_{L_4,1}$, and $G^{e'}_{L_4,2}$, respectively, by \citet{Pousse17}. Similar results for a moderate extent of the secondary's eccentricity may be found in \citet{Bien1978} and \citet{Nesvorny02}. \citet{Lidov94} focused only on the computation of quasi-satellite orbits and provided numerical and analytical methods and results for orbits of  various  multiplicities and mass parameters in the 2D-ERTBP.

However, in Figs. \ref{2DE}b,c we show new families of asymmetric periodic orbits that we discovered. They are called $E^A_{41}$, $E^A_{42}$, and $E^A_{43}$ and belong to the high-eccentricity 2D-ERTBP regime. These families were generated by the new bifurcation point, $B_T^4$, which was discovered along the new $A_2$ family featured in P4. 

Moreover, we provide new isolated families that are not linked with bifurcation points. These possess asymmetric periodic orbits in the 2D-ERTBP and are called $E^A_{44}$, $E^A_{45}$, and $E^A_{46}$. 

Finally, we computed the short- and long-period asymmetric (tadpole) orbits, which emanate from the $L_4$ of the ERTBP. The family of short-period asymmetric orbits, called $E^A_S$, has two branches. One branch ends when the eccentricity of the massless body (dust particle) and the eccentricity of the minor primary (secondary) become equal to zero. The termination of this branch is the $L_4$ of the CRTBP (see Fig. \ref{aL4}c). The termination of the other branch of $E^A_S$ takes place when the orbit the secondary (asteroid or exoplanet) is highly eccentric. \citet{Henrard1970,Henrard1970a} showcased the complexity of the families starting from $L_4$ for co-orbital motion, \citet{Bien1978} investigated the case of $\mu=10^{-3}$ in the 2D-ERTBP, and \citet{Lohinger1993}, \citet{Erdi2007}, and \citet{liwi20} explored the stability regions around $L_4$ for various $\mu$ values in the ERTBP. A respective study for the CRTBP (see P4) was carried out by \cite{Pousse17} and \cite{liwi20}, while the complexity of branches around $L_4$ in the CRTBP (P4) was shown by \citet{deprit67}. 

Although we did not show new families in the 3D-ERTBP, we included a continuation of the known families for mass parameters $\mu\in [10^{-10},10^{-5}]$ in the appendix (see Figs. \ref{3DE} and \ref{3deexp}).

\section{Phase space visualisation}\label{maps}

We computed DS maps with the use of the de-trended fast Lyapunov indicator (DFLI; \citealt{voyatzis08}). We adopted a computational setup similar to that of P4 and integrated each initial condition for $t_{max}=2.5$ Myr. Dark (pale) colours correspond to long-term stability (irregularity), white denotes collisions or failure of the integration, and dashed cyan curves mark close encounters. Here, we identified more combinations of libration/rotation of the resonant angle $\theta$; the differences $\Delta\varpi$ and $\Delta\Omega$ are directly related to the periodic orbits shown in Sect. \ref{ertbp} and signify each region of stability with a capital letter denoting a particular behaviour. Examples of such evolutions are included in Appendix \ref{app3}.  

\begin{figure}[H]
$\begin{array}{c}
\includegraphics[width=0.89\columnwidth]{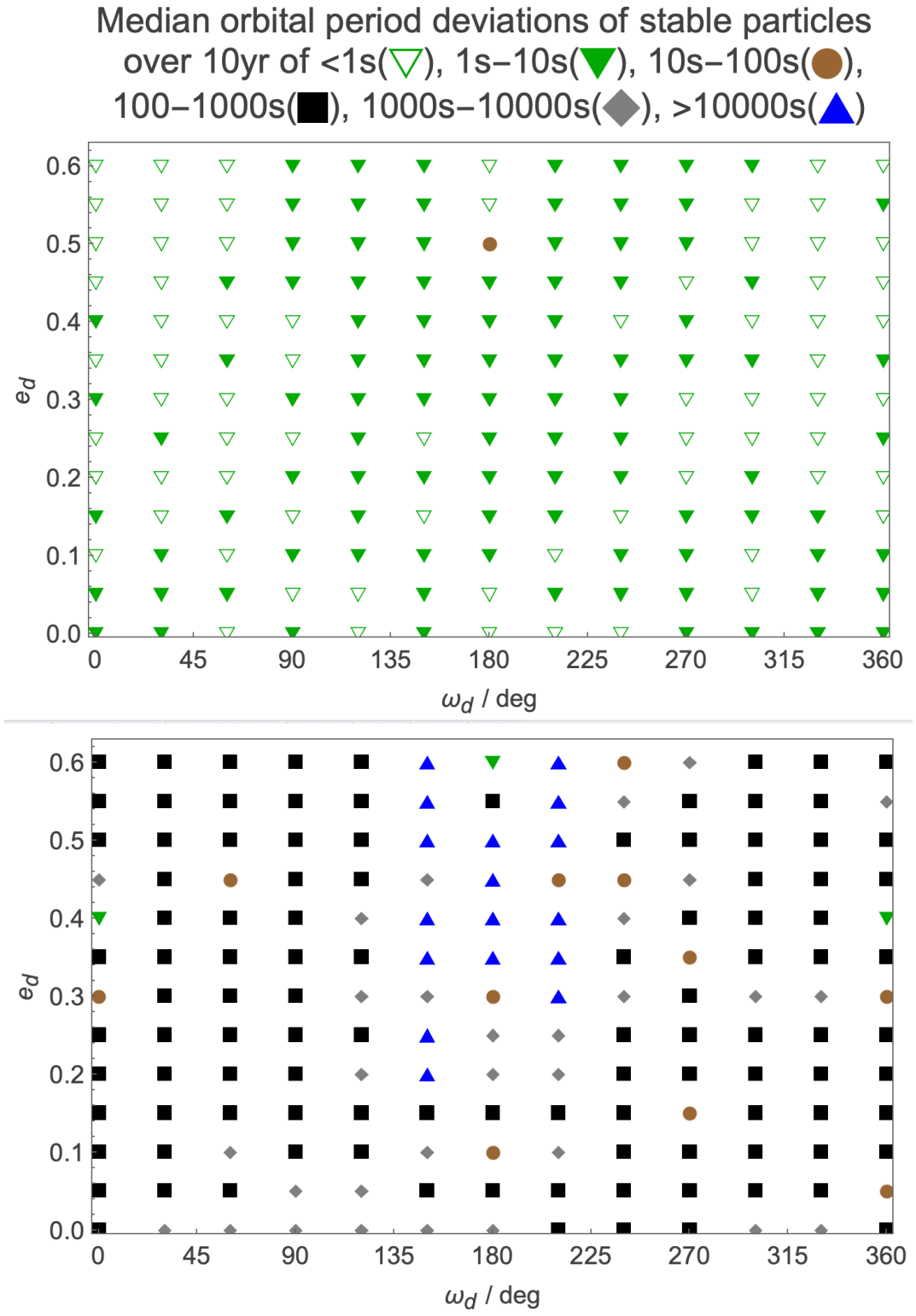}
\end{array}$\vspace{-0.25cm}
\caption{Simulation outcomes over a ten-year evolution on the $(\omega_{\rm d},e_{\rm d})$ plane guided by a stable planar periodic orbit with $e_{\rm a}=0.6$  of the $E_{Ia}$ family for Ceres (\textit{top}) and Mercury (\textit{bottom}).} \vspace{-0.25cm}
\label{sE1a_Bwe}
\end{figure}

\begin{figure}[H]
$\begin{array}{c}
\includegraphics[width=0.89\columnwidth]{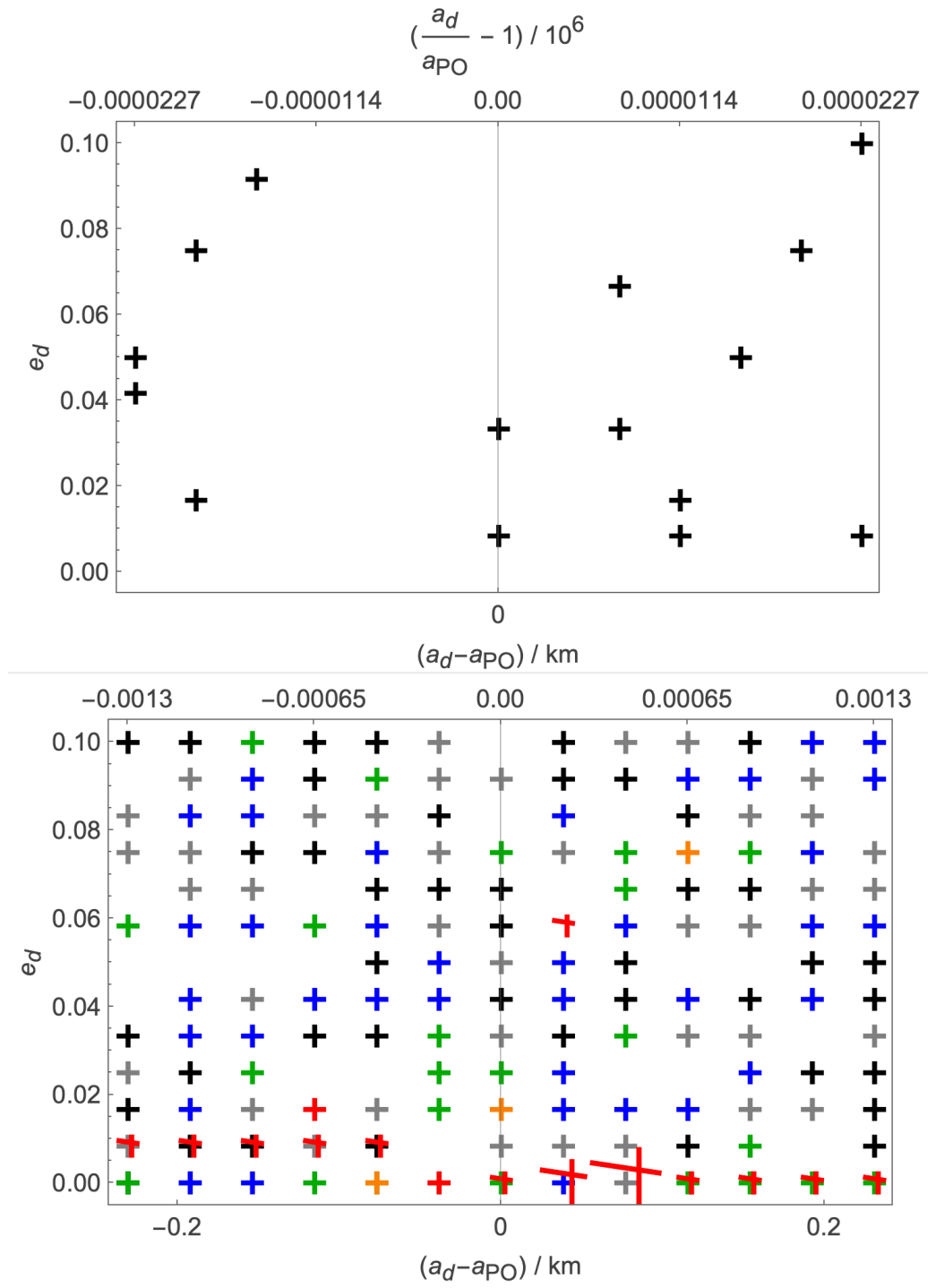}
\end{array}$\vspace{-0.25cm}
\caption{Collisions on the $(a_{\rm d},e_{\rm d})$ plane guided by a stable planar periodic orbit with $e_{\rm a}$$=$$0.6$  of the  $E_{Ia}$ family for Ceres (\textit{top}) and  Mercury (\textit{bottom}). The symbol colours and shapes indicate the number of the 18 test particles for each initial condition that collided with the planet: $(\{1,2,3,4,5,6,>6\}$ $= \{{\rm Black, Gray, Blue, Green, Orange, Red, Red \,Slanted}\})$.} \vspace{-0.25cm}
\label{cE1a_Bae}
\end{figure}

In Fig. \ref{2DE_TB} and for each secondary body we chose a planar stable symmetric periodic orbit with $e_{\rm a}=0.6$ from the planar $E_{1a}$ family shown in Fig. \ref{2DE}a and visualised the phase space on the $(a_{\rm d}, e_{\rm d})$ and the $(\omega_{\rm d}, e_{\rm d})$ planes. Symbol $\rm A$ stands for libration of both the resonant angle $\theta$ and the apsidal difference $\Delta\varpi$ and designates an area of e-resonance (MMR shown in Fig. \ref{Bottom_A}). Symbol $\rm B$ stands for the libration of $\theta$ and the rotation of $\Delta\varpi$ (see Fig. \ref{Bottom_B}). In these regions, the dust's eccentricity experienced high amplitude oscillations leading to a collision with the WD, since the phases were not protected via an MMR. Symbol $\rm C$ is chosen to specifically report areas linked with planar asymmetric periodic orbits. All angles librate (likewise in $\rm A$), yet they do so around angles different than $0^{\circ}$ or $180^{\circ}$ (see Fig. \ref{Bottom_C}).  Symbol $\rm D$ delineates regions where all angles rotate.

Next, we selected the specific 3D periodic orbits shown in Fig. \ref{3DE} and showcased phase space portraits on various combinations of orbital elements depending on the configuration of the spatial family. More particularly, the $F_a$ family guided the DS maps in Fig. \ref{3DE_Di10}, where the  mutual inclination was set to $\Delta i=10^{\circ}$, while the $F_p$ family provided the initial conditions with $\Delta i=45^{\circ}$ in the DS maps of Fig. \ref{3DE_Di45}. When exploring the spatial vicinities, we additionally computed grids on the $(e_{\rm d}, i_{\rm d})$, $(\omega_{\rm d}, \Omega_{\rm d})$ and $(\Omega_{\rm d}, M_{\rm d})$ planes. These DS maps showcase the effect of the variation of the apsidal and nodal difference. As reported by the initial conditions above each DS map, the semi-major axis of the dust particle varies along the planar and spatial families in the ERTBP (however, the semi-major axis ratio and hence the mean motion ratio is fixed to 1/1 MMR) as the mass parameter increases. Therefore, the regions of stability in relation to the angles shrink.

In Fig. \ref{3DE_Di10}, symbol $\rm A$ stands for libration of all angles ($\theta$, $\Delta\varpi,$ and $\Delta\Omega$) and refers to an area of both e-resonance and i-resonance (MMR in 3D periodic orbits; see also Fig. \ref{Fa_A}). In areas labelled by $\rm B,$ only $\theta$ librates (no MMR). Symbol $\rm C$ is linked to an MMR of the planar asymmetric periodic orbits (only e-resonance, no i-resonance; see Fig. \ref{Fp_C}). Symbol $\rm D$ is associated with apsidal difference oscillation ($\theta$ and $\Delta\Omega$ rotate), and symbol $\rm E$ determines long-term irregularity as all angles rotate therein.

\begin{figure}[H]
$\begin{array}{c}
\includegraphics[width=0.9\columnwidth]{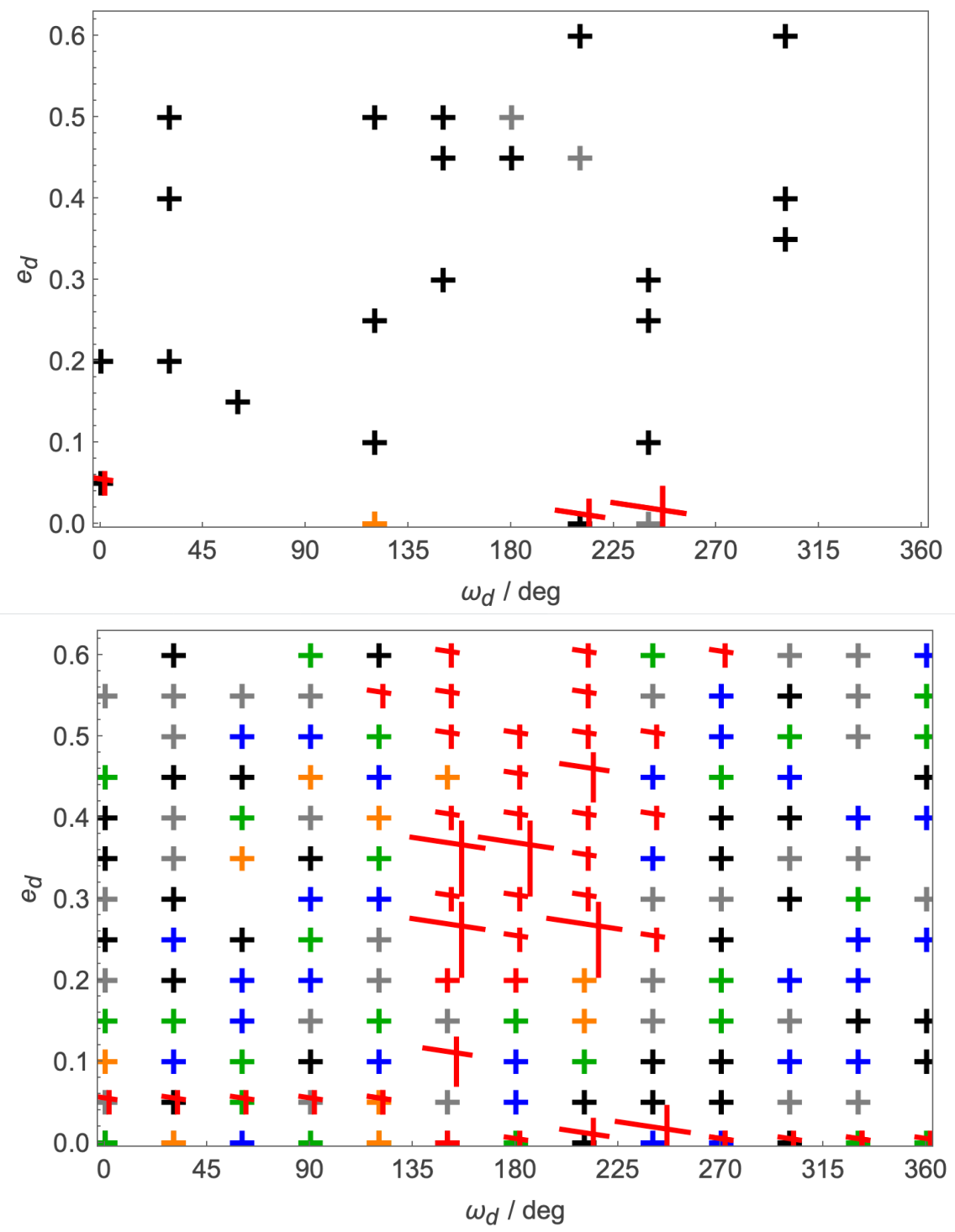}
\end{array}$\vspace{-0.25cm}
\caption{Collisions on  $(\omega_{\rm d},e_{\rm d})$ plane guided by a stable planar periodic orbit with $e_{\rm a}$$=$$0.6$  of  $E_{Ia}$ family for Ceres (\textit{top}) and  Mercury (\textit{bottom}). The symbol colours and shapes indicate the number of the 18 test particles for each initial condition that collided with the planet: $(\{1,2,3,4,5,6,>6\}$ $= \{{\rm Black, Gray, Blue, Green, Orange, Red, Red \,Slanted}\})$.}
\label{cE1a_Bwe}
\end{figure}

In Fig. \ref{3DE_Di45} we present the vicinity of an unstable 3D symmetric periodic orbit. Its immense irregular effect on long time spans is detected in high $\mu$ values (pale and white colours). Symbols $\rm B$,  $\rm C,$ and $\rm D$ denote a similar behaviour to those in Fig. \ref{3DE_Di10}. However, when the dust's inclination value increased ($\Delta i=45^{\circ}$), one more combination was exhibited with regard to the evolution of the angles. More particularly, in areas tagged with $\rm E$, a nodal difference oscillation (with $\theta$ and $\Delta\varpi$ rotating) is observed (see Fig. \ref{Fp_E}), while around $\rm F$ all angles rotate.

\section{N-body simulations}\label{sims}

As in P1, P2, P3, and P4, we performed $N$-body simulations to complement the computation of periodic orbits and stability maps. An objective of performing these simulations was to compute orbital period deviations of the secondary and collisional activity in the close vicinity of the white dwarf. We now reiterate from P4 why computing these quantities is important.

\begin{figure}[H]
$\begin{array}{c}
\includegraphics[width=0.9\columnwidth]{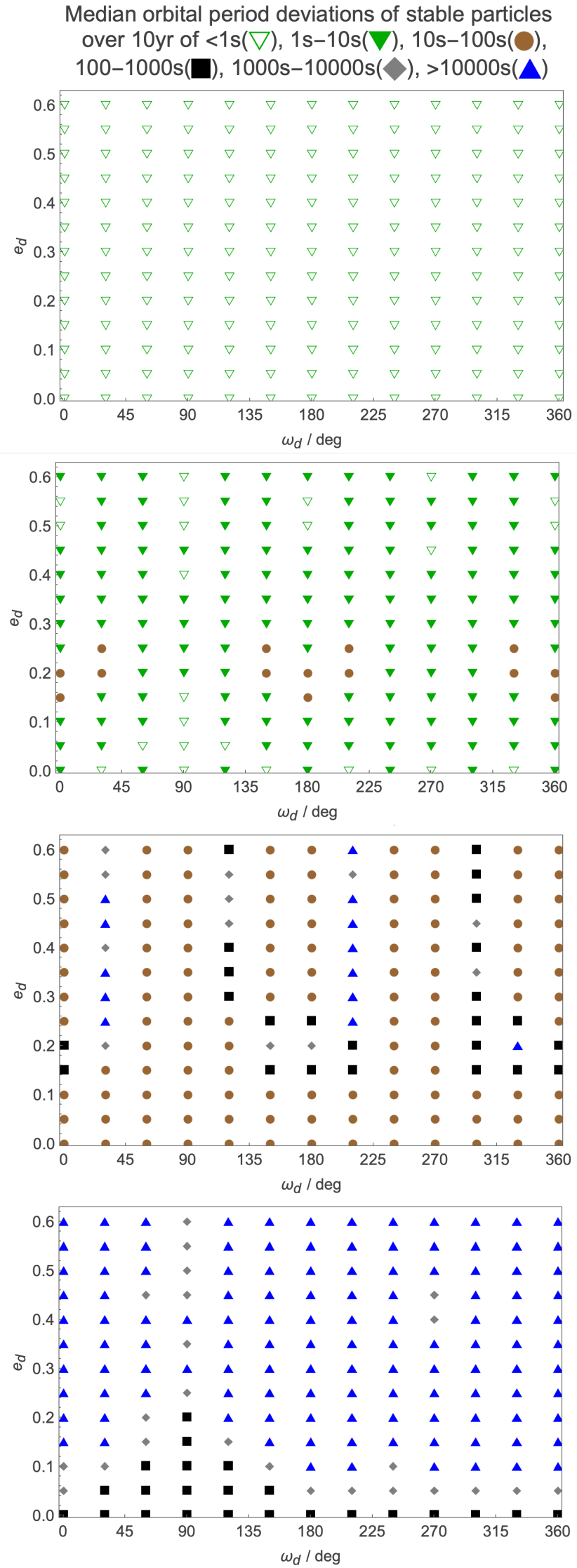}
\end{array}$\vspace{-0.25cm}
\caption{Simulation outcomes over a ten-year evolution on the $(\omega_{\rm d},e_{\rm d})$ plane guided by a stable 3D periodic orbit with $\Delta i=10^{\circ}$  of the $F_{a}$ family for Ceres, Mercury, Earth, and Neptune (first to fourth rows).}
\label{sFa_we}
\end{figure}

\begin{figure}[H]
$\begin{array}{c}
\includegraphics[width=0.9\columnwidth]{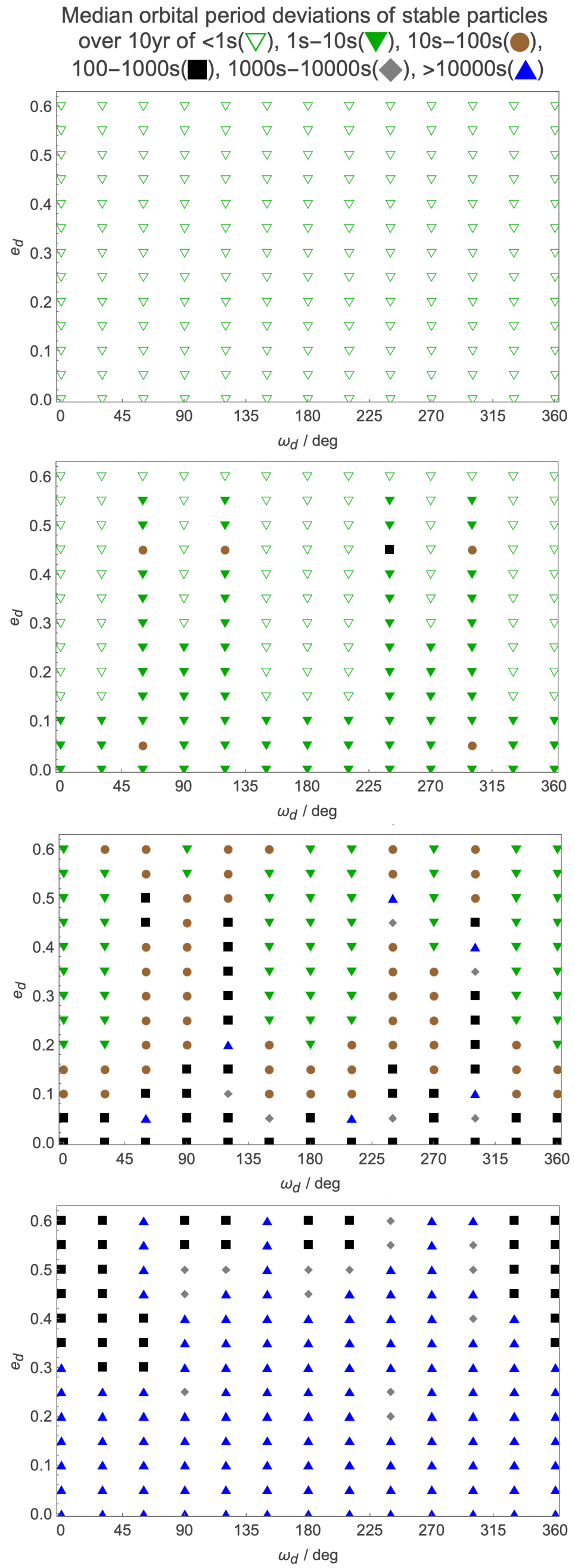}
\end{array}$\vspace{-0.25cm}
\caption{Simulation outcomes over a ten-year evolution on the $(\omega_{\rm d},e_{\rm d})$ plane guided by a stable 3D periodic orbit with $\Delta i=45^{\circ}$  of the $F_p$ family for Ceres, Mercury, Earth, and Neptune (first to fourth rows).}
\label{sFp_we}
\end{figure}

Orbital period deviations are reflective of transit timing variations, which are directly observable quantities. Observers see debris crossing in front of a white dwarf through photometric transits. However, after some orbits, the crossing time may differ, indicating a change in orbital period or structure. This orbital shift may be induced by an unseen planet, and the more massive the planet, the larger the orbital period change \citep[e.g.][]{gurr17}.

Photometric transits of debris around white dwarfs are quickly time-varying (on timescales of weeks to years) and feature-rich (with shallow and sharp features in the transit curves), indicating a high degree of dynamical activity. One potential manner in which dust is created -- thereby changing the shape of the transit curve -- is through collisions with the secondary. Although the planet may remain hidden from view, the production of dust is clearly seen in some instances \citep{vander15,vander20,vander21,farihi22,bhat25,guid25,hermes25,korth26}.

Our simulation setup here is equivalent to that in P4 except that here our initial conditions are motivated by the families in the 2D-ERTBP and 3D-ERTBP instead of the families in the 2D-CRTBP and 3D-CRTBP. The families we investigated here are $E_{Ia}$, $F_a$, and $F_p$, since they are the only ones possessing stable orbits with $max(e_{\rm d})=0.6$ in tandem with $max(e_{\rm a})=0.6$.

\begin{figure}[H]
$\begin{array}{c}
\includegraphics[width=0.9\columnwidth]{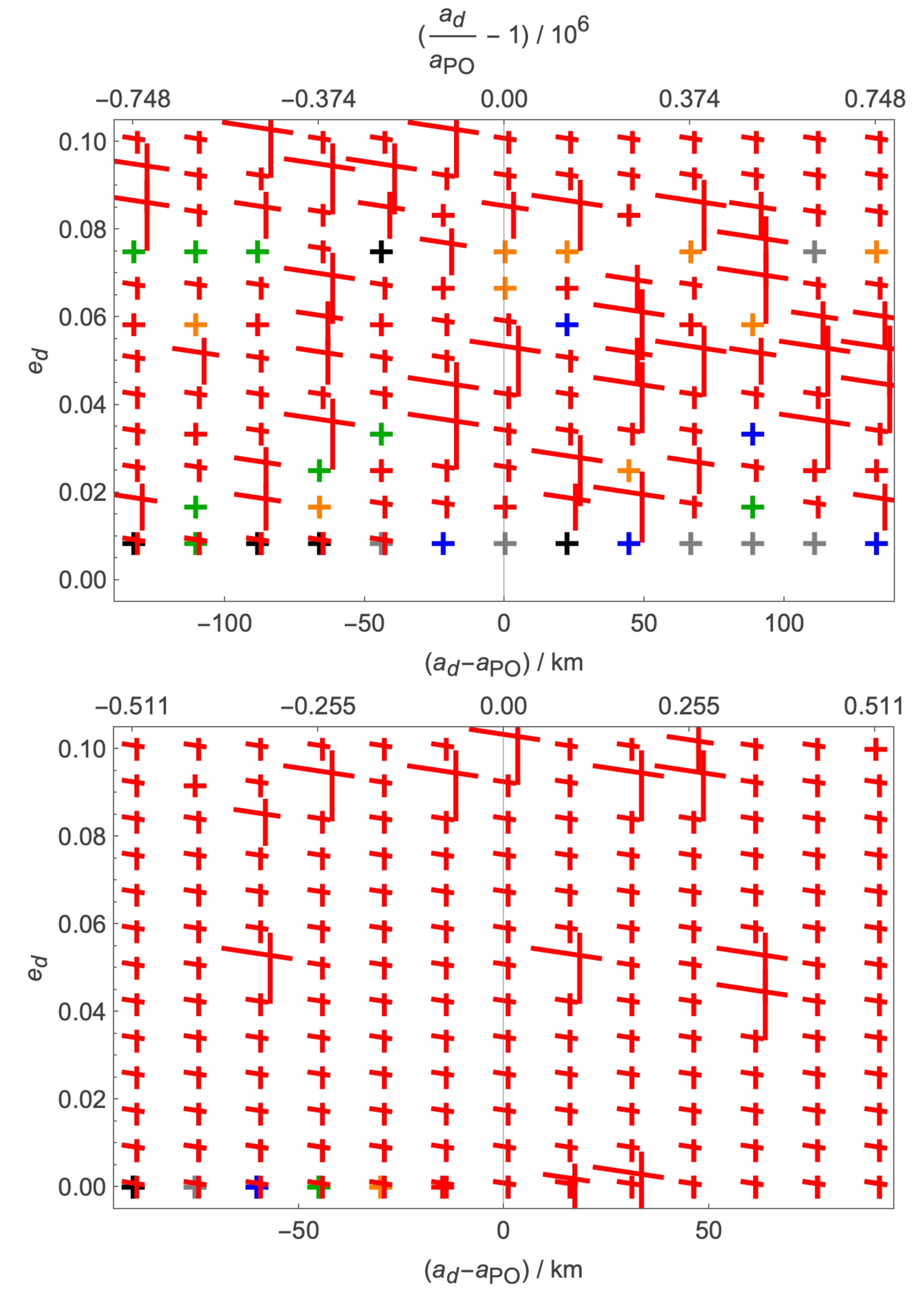}\\
\end{array}$\vspace{-0.25cm}
\caption{Collisions on the $(a_d,e_d)$  plane for Neptune and dust around a periodic orbit with $\Delta i=10^{\circ}$ of the $F_{a}$ family (\textit{top}) and $\Delta i=45^{\circ}$ of the $F_{p}$ family (\textit{bottom}). The symbol colours and shapes indicate the number of the 18 test particles for each initial condition that collided with the planet: $(\{1,2,3,4,5,6,>6\}$ $= \{{\rm Black, Gray, Blue, Green, Orange, Red, Red \,Slanted}\})$.}\vspace{-0.25cm}
\label{cFa}
\end{figure}

To briefly reiterate our simulation setup, we used the {\tt IAS15} integrator in {\tt REBOUND} \citep{rein2012,rein2015} with a minimum time step of 63.1 sec and suitably scaled lengths (including distances and radii) corresponding to $a_{\rm a} = 0.012$~au. The mass of the WD is 0.6 $M_\odot,$ and the collision radius is the Roche radius. The canonical WD Roche radius resides at 1 $R_\odot$, or about 0.0047 au. We integrated systems corresponding to all four mass parameters. We ran our simulations for ten years, as the first WD with observable transiting debris was discovered in 2015 \citep{vander15}, and we have been observing at least that system for about a decade. We output both orbital period deviations of the dust particles and the number of collisions with the secondary that occurs. Each simulation was run with 18 test particles that were initially uniformly distributed in mean anomaly; for each corresponding plot, the black, grey, blue, green, orange, red, and slanted red symbols indicate one, two, three, four, five, six, or more than six particles (out of 18) that collided with the secondary.

In the following, figures featuring mean orbital period deviations are applicable to observations because the orbital period is a direct observable. The figures showcasing collisional outcomes are more indirectly related to observables -- these collisions generate dust, which produces transit dips, which are observable. However, the locations of the dust generation is degenerate because the dust can spread and be produced in different amounts depending on the physical characteristics of the collisions and the amount of gas that is sublimated.

\begin{figure}
$\begin{array}{c}
\includegraphics[width=0.89\columnwidth]{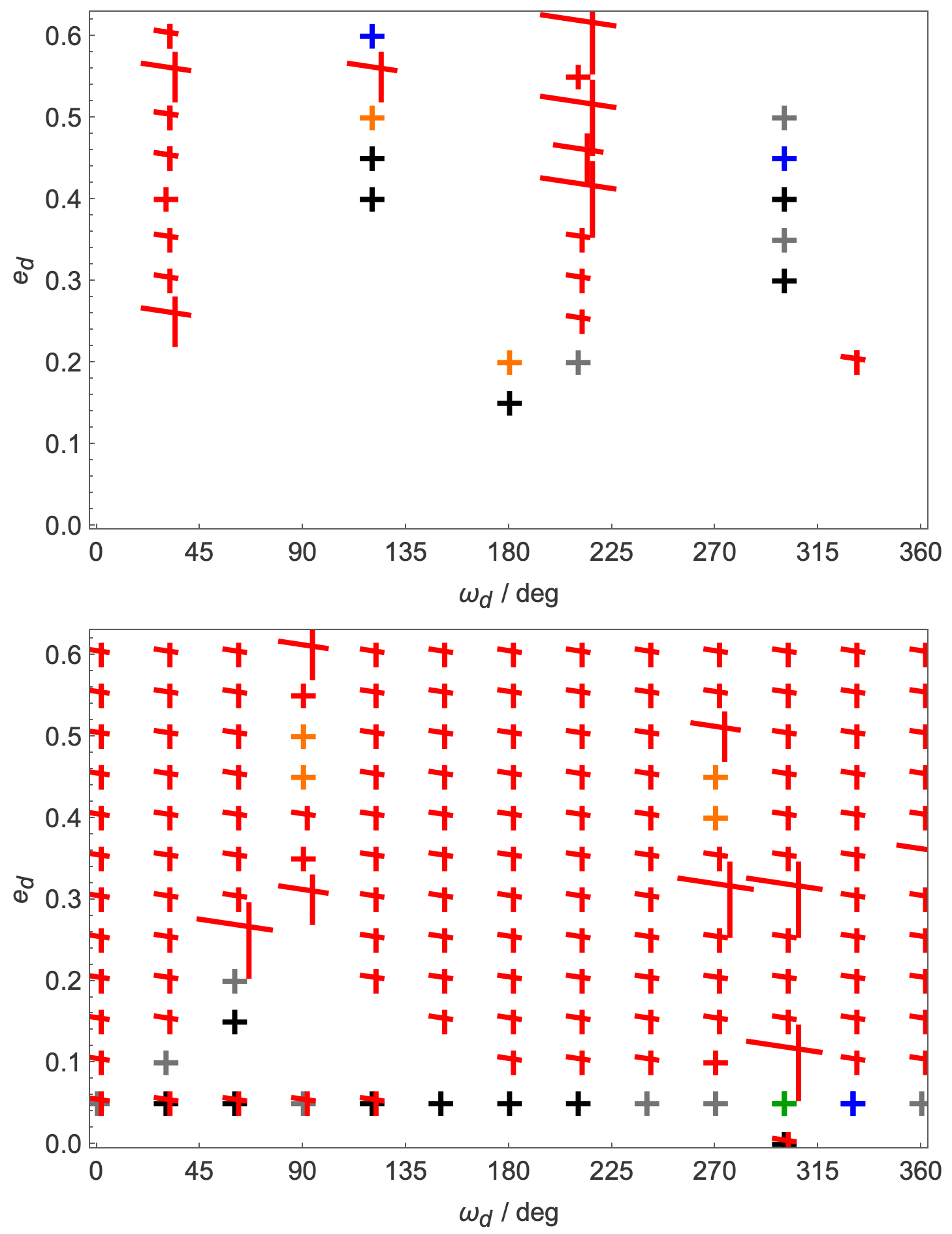}\\
\end{array}$\vspace{-0.25cm}
\caption{Collisions on the $(\omega_d,e_d)$  plane for Earth (\textit{top}) and Neptune (\textit{bottom}) and dust around a periodic orbit with $\Delta i=10^{\circ}$ of the $F_{a}$ family. The symbol colours and shapes indicate the number of the 18 test particles for each initial condition that collided with the planet: $(\{1,2,3,4,5,6,>6\}$ $= \{{\rm Black, Gray, Blue, Green, Orange, Red, Red \,Slanted}\})$.}\vspace{-0.25cm}
\label{cFa_we}
\end{figure}

\begin{figure}
$\begin{array}{c}
\includegraphics[width=0.89\columnwidth]{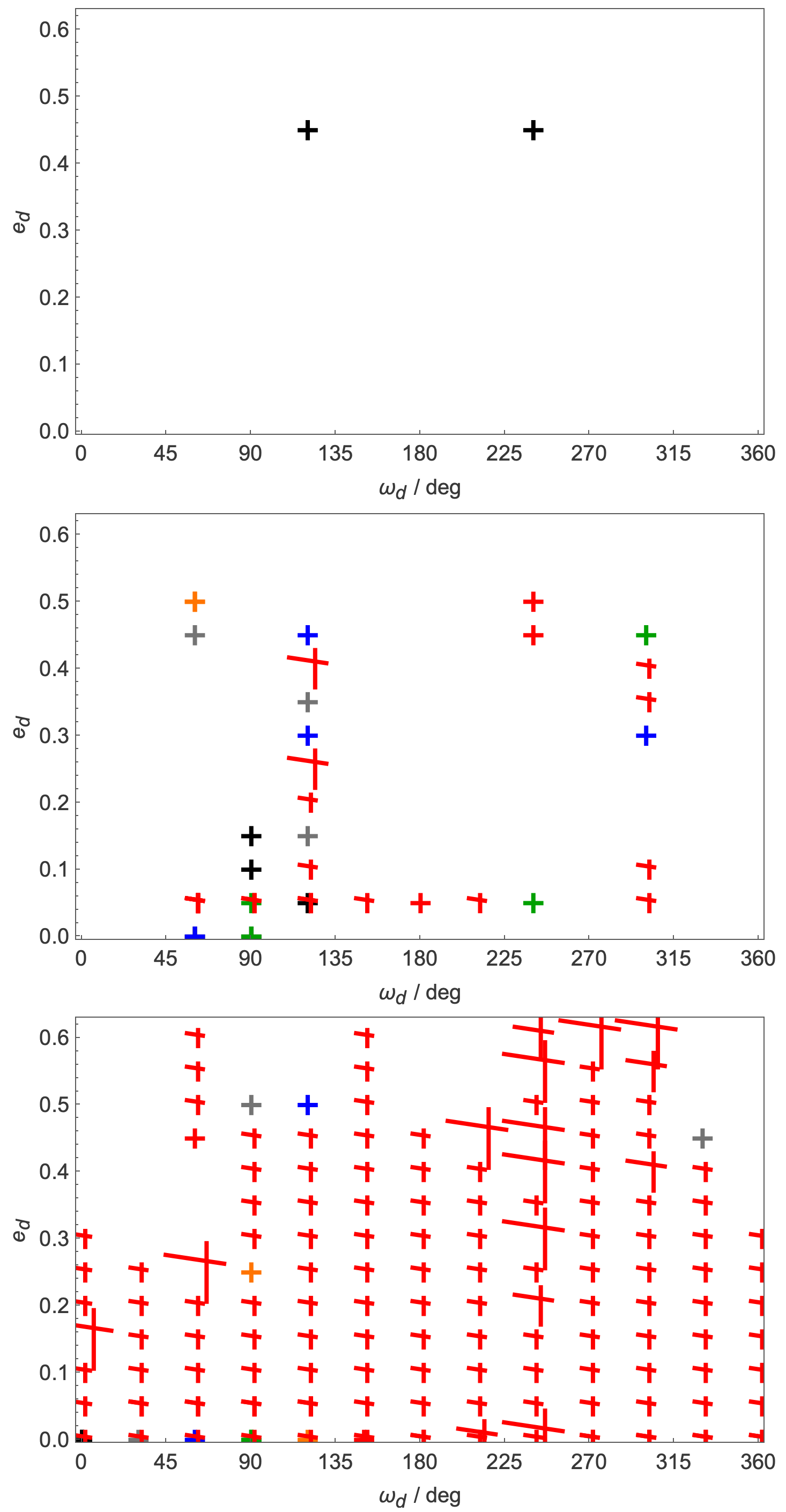}\\
\end{array}$\vspace{-0.25cm}
\caption{Collisions on the $(\omega_d,e_d)$  plane for Mercury (\textit{top}), Earth (\textit{middle}), and Neptune (\textit{bottom}) and dust around a periodic orbit with $\Delta i=45^{\circ}$ of the $F_{p}$ family. The symbol colours and shapes indicate the number of the 18 test particles for each initial condition that collided with the planet: $(\{1,2,3,4,5,6,>6\}$ $= \{{\rm Black, Gray, Blue, Green, Orange, Red, Red \,Slanted}\})$.}\vspace{-0.25cm}
\label{cFp_we}
\end{figure}

We plot only the most interesting outcomes, which are those that showcased significant variation across the explored parameter space. Our results are presented in Figs. \ref{sE1a_Bwe}-\ref{cE1a_Bwe} for family $E_{Ia}$ (for Ceres-like and Mercury-like secondaries only), Figs. \ref{sFa_we}, \ref{cFa}, and \ref{cFa_we} for family $F_a$, and Figs. \ref{sFp_we} and \ref{cFp_we} for family $F_p$. 

A comparison of the orbital period deviations of all three families results in a marked contrast for $E_{Ia}$: the Mercury-like secondary generates much larger (by orders of magnitude) orbital period deviations than for $F_a$ and $F_p$. The difference is a significant indicator of non-coplanarity in the system. Between the orbital period deviations for $F_a$ and $F_p$ (comparing Figs. \ref{sFa_we}-\ref{sFp_we}), the difference is not as great, although the deviations are smaller for $F_p$ than for $F_a$.

The collisional activity follows suit: it is greatest for $E_{Ia}$, followed by $F_a$ and then $F_p$. This activity shows some marked trends as a function of $\omega_d$. In particular, certain values of $\omega_d$ prompt spikes in activity. In Fig. \ref{cE1a_Bwe} this value is around $180^{\circ}$, whereas in Figs. \ref{cFa_we} and \ref{cFp_we} the spikes occur at different values, and sometimes only above a particular value of $e_d$.\vspace{-0.2cm}

\section{Discussion}\label{dis}

White dwarf planetary systems continue to represent a basis on which to discover new periodic-orbit families. These systems also hide orbiting objects whose existence can at least be hinted at thanks to these families and using dynamical stability maps and $N$-body simulations. We do not claim that our results can precisely pinpoint the location and mass of hidden planets, but rather they reduce the degeneracy in the parameter space and serve as a useful starting point for detailed simulations on an individual system basis.

A comparison of the results of this work and P4 showcase how the results from the ERTBP are sufficiently different from those of the CRTBP to enable predictions for orbital parameters of secondary objects given adequately detailed observations. Overall, the DS maps shown here are more feature-rich and asymmetric than those in P4, which aids in breaking degeneracies when inferring secondary orbital parameters. For the $N$-body simulation results, Fig. \ref{sE1a_Bwe} highlights the most significant difference; eccentric Ceres-like secondaries can produce deviations of the order of one to ten seconds, whereas the circular equivalent nearly always generates deviations of under one~second.

The differences between the 2D-ERTBP and 3D-ERTBP are also important. Objects scattered towards a white dwarf do not have a preferential 3D orientation unless the scatterer is small and the initial configuration is near-coplanar \citep{verasgeo21}. Hence, planets and large asteroids may be present but hidden from our line of sight in transit observations of the dust. Further, the observed dust may not be associated with the orbiting secondary, but may be remnants of a previous break-up event. With 25--50\% of white dwarfs containing planetary remnants in their atmospheres \citep{zuck10,koester14,ould24}, such scattering and tidal break-up events may be common and overlap with one another, albeit at different inclinations.

\section{Conclusions}\label{concl}

The purpose of our study is threefold: 1) to provide a methodological presentation of the continuation of the families of symmetric and asymmetric periodic orbits in the ERTBP, which underlines the complexity of the phase space in the ERTBP and shows that the semi-major axes, along with the orbital angles of the bodies, play a crucial role in the evolution of real systems; 2) to generate a global view of the phase space around WDs along with their (hidden or partially disrupted) secondaries and dust particles; and 3) to rigorously yet straightforwardly compute observable orbital period deviations via $N$-body simulations that adhere to the periodic orbit families.
In this work, we computed families of periodic orbits in the 1/1 MMR in the 2D-ERTBP and the 3D-ERTBP. New families of asymmetric periodic orbits, called $E^A_{41}$, $E^A_{42}$, and $E^A_{43}$ in the high-eccentricity 2D-ERTBP regime were discovered. They were generated by the new bifurcation point, $B_T^4$, along the new $A_2$ family, showcased by \citet{ave26a}. Moreover, new isolated families of asymmetric periodic orbits, namely $E^A_{44}$, $E^A_{45}$, and $E^A_{46}$ were discovered, while families of short- and long- period asymmetric orbits starting from $L_4$, called $E^A_S$  and $E^A_L$, were computed. We also identified the link between specific islands of stability and debris disc maintenance through $N$-body simulations with i) isolated families of asymmetric periodic orbits and ii) families of stable asymmetric periodic orbits starting from $L_4$.

The periodic-orbit seeded $N$-body simulations revealed that over ten years Ceres-like secondaries can produce deviations of the order of one to ten seconds, whereas the circular equivalent nearly always generates deviations of under one~second. (see \citealt{ave26a}). In general, the orbital period deviations of the dust are higher than in the CRTBP; in some architectures, Mercury-mass planets can regularly generate deviations exceeding $\sim$100 s, which potentially aids observational inferences of hidden asteroids or planets orbiting WDs.

With Gaia Data Release 4, which is scheduled for release before 2027, the number of transiting white dwarf planetary systems whose secondaries are in a transitional eccentric phase before circularisation might increase. The results in this paper provide a starting basis on which to perform a detailed analysis for individual systems.

\begin{acknowledgements} We thank the anonymous reviewer for their insightful comments that improved the clarity of the presentation. KIA gratefully acknowledges the support provided by the Digital Governance Unit of Aristotle University of Thessaloniki (AUTh); Results have been produced using the AUTh HPC Infrastructure and Resources. 
\end{acknowledgements}

\bibliographystyle{aa}
\bibliography{11}

@ARTICLE{zuck10,
       author = {{Zuckerman}, B. and {Melis}, C. and {Klein}, B. and {Koester}, D. and {Jura}, M.},
        title = "{Ancient Planetary Systems are Orbiting a Large Fraction of White Dwarf Stars}",
      journal = {\apj},
         year = 2010,
        month = oct,
       volume = {722},
       number = {1},
        pages = {725-736},
          doi = {10.1088/0004-637X/722/1/725},
archivePrefix = {arXiv},
       eprint = {1007.2252},
 primaryClass = {astro-ph.SR},
       adsurl = {https://ui.adsabs.harvard.edu/abs/2010ApJ...722..725Z}
}

@ARTICLE{koester14,
       author = {{Koester}, D. and {G{\"a}nsicke}, B.~T. and {Farihi}, J.},
        title = "{The frequency of planetary debris around young white dwarfs}",
      journal = {\aap},
         year = 2014,
        month = jun,
       volume = {566},
          eid = {A34},
        pages = {A34},
          doi = {10.1051/0004-6361/201423691},
archivePrefix = {arXiv},
       eprint = {1404.2617},
 primaryClass = {astro-ph.SR},
       adsurl = {https://ui.adsabs.harvard.edu/abs/2014A&A...566A..34K}
}

@ARTICLE{ould24,
       author = {{Ould Rouis}, Lou Baya and {Hermes}, J.~J. and {G{\"a}nsicke}, Boris T. and {Sahu}, Snehalata and {Koester}, Detlev and {Tremblay}, P.-E. and {Veras}, Dimitri and {Farihi}, Jay and {Heintz}, Tyler M. and {Gentile Fusillo}, Nicola Pietro and {Redfield}, Seth},
        title = "{Constraints on Remnant Planetary Systems as a Function of Main-sequence Mass with HST/COS}",
      journal = {\apj},
         year = 2024,
        month = dec,
       volume = {976},
       number = {2},
          eid = {156},
        pages = {156},
          doi = {10.3847/1538-4357/ad86bb},
archivePrefix = {arXiv},
       eprint = {2410.06335},
 primaryClass = {astro-ph.SR},
       adsurl = {https://ui.adsabs.harvard.edu/abs/2024ApJ...976..156O}
}

@ARTICLE{verasgeo21,
       author = {{Veras}, Dimitri and {Georgakarakos}, Nikolaos and {Mustill}, Alexander J. and {Malamud}, Uri and {Cunningham}, Tim and {Dobbs-Dixon}, Ian},
        title = "{The entry geometry and velocity of planetary debris into the Roche sphere of a white dwarf}",
      journal = {\mnras},
         year = 2021,
        month = sep,
       volume = {506},
       number = {1},
        pages = {1148-1164},
          doi = {10.1093/mnras/stab1667},
archivePrefix = {arXiv},
       eprint = {2106.04601},
 primaryClass = {astro-ph.EP},
       adsurl = {https://ui.adsabs.harvard.edu/abs/2021MNRAS.506.1148V}
}

@ARTICLE{vanderburg20,
       author = {{Vanderburg}, Andrew and {Rappaport}, Saul A. and {Xu}, Siyi and {Crossfield}, Ian J.~M. and {Becker}, Juliette C. and {Gary}, Bruce and {Murgas}, Felipe and {Blouin}, Simon and {Kaye}, Thomas G. and {Palle}, Enric and {Melis}, Carl and {Morris}, Brett M. and {Kreidberg}, Laura and {Gorjian}, Varoujan and {Morley}, Caroline V. and {Mann}, Andrew W. and {Parviainen}, Hannu and {Pearce}, Logan A. and {Newton}, Elisabeth R. and {Carrillo}, Andreia and {Zuckerman}, Ben and {Nelson}, Lorne and {Zeimann}, Greg and {Brown}, Warren R. and {Tronsgaard}, Ren{\'e} and {Klein}, Beth and {Ricker}, George R. and {Vanderspek}, Roland K. and {Latham}, David W. and {Seager}, Sara and {Winn}, Joshua N. and {Jenkins}, Jon M. and {Adams}, Fred C. and {Benneke}, Bj{\"o}rn and {Berardo}, David and {Buchhave}, Lars A. and {Caldwell}, Douglas A. and {Christiansen}, Jessie L. and {Collins}, Karen A. and {Col{\'o}n}, Knicole D. and {Daylan}, Tansu and {Doty}, John and {Doyle}, Alexandra E. and {Dragomir}, Diana and {Dressing}, Courtney and {Dufour}, Patrick and {Fukui}, Akihiko and {Glidden}, Ana and {Guerrero}, Natalia M. and {Guo}, Xueying and {Heng}, Kevin and {Henriksen}, Andreea I. and {Huang}, Chelsea X. and {Kaltenegger}, Lisa and {Kane}, Stephen R. and {Lewis}, John A. and {Lissauer}, Jack J. and {Morales}, Farisa and {Narita}, Norio and {Pepper}, Joshua and {Rose}, Mark E. and {Smith}, Jeffrey C. and {Stassun}, Keivan G. and {Yu}, Liang},
        title = "{A giant planet candidate transiting a white dwarf}",
      journal = {\nat},
         year = 2020,
        month = sep,
       volume = {585},
       number = {7825},
        pages = {363-367},
          doi = {10.1038/s41586-020-2713-y},
archivePrefix = {arXiv},
       eprint = {2009.07282},
 primaryClass = {astro-ph.EP},
       adsurl = {https://ui.adsabs.harvard.edu/abs/2020Natur.585..363V}
}

@ARTICLE{gans19,
       author = {{G{\"a}nsicke}, Boris T. and {Schreiber}, Matthias R. and {Toloza}, Odette and {Gentile Fusillo}, Nicola P. and {Koester}, Detlev and {Manser}, Christopher J.},
        title = "{Accretion of a giant planet onto a white dwarf star}",
      journal = {\nat},
         year = 2019,
        month = dec,
       volume = {576},
       number = {7785},
        pages = {61-64},
          doi = {10.1038/s41586-019-1789-8},
archivePrefix = {arXiv},
       eprint = {1912.01611},
 primaryClass = {astro-ph.EP},
       adsurl = {https://ui.adsabs.harvard.edu/abs/2019Natur.576...61G}
}

@ARTICLE{korth26,
       author = {{Korth}, J. and {Mustill}, A.~J. and {Parviainen}, H. and {Villaver}, E. and {Kuehne}, J.~W. and {B{\'e}jar}, V.~J.~S. and {Hayashi}, Y. and {Abreu Garc{\'\i}a}, N. and {Kagetani}, T. and {Kawauchi}, K. and {Livingston}, L. and {Mori}, M. and {Morello}, G. and {Watanabe}, N. and {Fukuda}, I. and {Ikuta}, K. and {Bonilla-Mariana}, I. and {Esparza-Borges}, E. and {Fern{\'a}ndez-Rodr{\'\i}guez}, G. and {Fukui}, A. and {Gerald{\'\i}a-Gonz{\'a}lez}, S. and {Gonz{\'a}lez-Rodr{\'\i}guez}, J. and {Isogai}, K. and {Narita}, N. and {Palle}, E. and {Pel{\'a}ez-Torres}, A. and {S{\'a}nchez-Benavente}, M.},
        title = "{WD 1054-226 revisited: A stable transiting debris system}",
      journal = {A\&A},
         year = 2026,
        month = apr,
       volume = {709},
          eid = {A27},
        pages = {A27},
          doi = {10.1051/0004-6361/202558102},
archivePrefix = {arXiv},
       eprint = {2603.07677},
 primaryClass = {astro-ph.EP},
       adsurl = {https://ui.adsabs.harvard.edu/abs/2026A&A...709A..27K}
}

@ARTICLE{bhat25,
       author = {{Bhattacharjee}, Soumyadeep and {Vanderbosch}, Zachary P. and {Hollands}, Mark A. and {Tremblay}, Pier-Emmanuel and {Xu}, Siyi and {Guidry}, Joseph A. and {Hermes}, J.~J. and {Caiazzo}, Ilaria and {Rodriguez}, Antonio C. and {van Roestel}, Jan and {El-Badry}, Kareem and {Drake}, Andrew J. and {Roulston}, Benjamin R. and {Riddle}, Reed and {Ben Rusholme} and {Groom}, Steven L. and {Smith}, Roger and {Toloza}, Odette},
        title = "{A ZTF Search for Circumstellar Debris Transits in White Dwarfs: Six New Candidates, One with Gas Disk Emission, Identified in a Novel Metric Space}",
      journal = {PASP},
         year = 2025,
        month = jul,
       volume = {137},
       number = {7},
          eid = {074202},
        pages = {074202},
          doi = {10.1088/1538-3873/ade0ea},
archivePrefix = {arXiv},
       eprint = {2502.05502},
 primaryClass = {astro-ph.SR},
       adsurl = {https://ui.adsabs.harvard.edu/abs/2025PASP..137g4202B}
}

@ARTICLE{hermes25,
       author = {{Hermes}, J.~J. and {Guidry}, Joseph A. and {Vanderbosch}, Zachary P. and {Badenas-Agusti}, Mariona and {Xu}, Siyi and {Kao}, Malia L. and {Rodriguez}, Antonio C. and {Hawkins}, Keith},
        title = "{Sporadic Dips from Extended Debris Transiting the Metal-rich White Dwarf SBSS 1232+563}",
      journal = {ApJ},
         year = 2025,
        month = feb,
       volume = {980},
       number = {1},
          eid = {56},
        pages = {56},
          doi = {10.3847/1538-4357/ada5fd},
archivePrefix = {arXiv},
       eprint = {2501.02050},
 primaryClass = {astro-ph.SR},
       adsurl = {https://ui.adsabs.harvard.edu/abs/2025ApJ...980...56H}
}

@ARTICLE{guid25,
       author = {{Guidry}, Joseph A. and {Vanderbosch}, Zachary P. and {Hermes}, J.~J. and {Veras}, Dimitri and {Hollands}, Mark A. and {Bhattacharjee}, Soumyadeep and {Caiazzo}, Ilaria and {El-Badry}, Kareem and {Kao}, Malia L. and {Ould Rouis}, Lou Baya and {Rodriguez}, Antonio C. and {van Roestel}, Jan},
        title = "{Transiting Planetary Debris near the Roche Limit of a White Dwarf on a 4.97 hr Orbit{\textemdash}and its Vanishing}",
      journal = {ApJ},
         year = 2025,
        month = oct,
       volume = {992},
       number = {2},
          eid = {167},
        pages = {167},
          doi = {10.3847/1538-4357/adfecb},
archivePrefix = {arXiv},
       eprint = {2508.18348},
 primaryClass = {astro-ph.SR},
       adsurl = {https://ui.adsabs.harvard.edu/abs/2025ApJ...992..167G}
}

@Article{hen,
  author   = {M. H{\'e}non},
  title    = {Vertical Stability of Periodic Orbits in the Restricted Problem. I. Equal Masses},
  journal  = {A\&A},
  year     = {1973},
  volume   = {28},
  pages    = {415},
}

@Book{cont,
  title     = {{Order and chaos in dynamical astronomy}},
  publisher = {Springer},
  year      = {2002},
  author    = {{Contopoulos}, G.},
  address   = {Berlin},
  booktitle = {Order and chaos in dynamical astronomy /George Contopoulos.~Berlin : Springer, c2002.~ (Astronomy and astrophys library) QB 351 .K58 2002},
}

@Article{voyatzis08,
  author   = {{Voyatzis}, George},
  title    = {{Chaos, Order, and Periodic Orbits in 3:1 Resonant Planetary Dynamics}},
  journal  = {\apj},
  year     = {2008},
  volume   = {675},
  number   = {1},
  pages    = {802-816},
  month    = mar,
  doi      = {10.1086/526513},
}

@Book{hen97,
  title     = {{Generating Families in the Restricted Three-Body Problem}},
  publisher = {Springer-Verlag Berlin Heidelberg},
  year      = {1997},
  author    = {{H{\'e}non}, Michel},
  volume    = {52},
  series    = {Lecture Notes in Physics Monographs},
  doi       = {10.1007/3-540-69650-4},
}

@Article{vmtg16,
  author  = {{Veras}, Dimitri and {Marsh}, Thomas R. and {G{\"a}nsicke}, Boris T.},
  title   = {{Dynamical mass and multiplicity constraints on co-orbital bodies around stars}},
  journal = {\mnras},
  year    = {2016},
  volume  = {461},
  number  = {2},
  pages   = {1413-1420},
  doi     = {10.1093/mnras/stw1324},
}

@Article{Lidov94,
  author  = {{Lidov}, M.~L. and {Vashkov'yak}, M.~A.},
  title   = {{On quasi-satellite orbits in a restricted elliptic three-body problem}},
  journal = {Astronomy Letters},
  year    = {1994},
  volume  = {20},
  number  = {5},
  pages   = {676-690},
  adsurl  = {http://adsabs.harvard.edu/abs/1994AstL...20..676L},
}

@Article{Pousse17,
  author  = {{Pousse}, A. and {Robutel}, P. and {Vienne}, A.},
  title   = {{On the co-orbital motion in the planar restricted three-body problem: the quasi-satellite motion revisited}},
  journal = {Celestial Mechanics and Dynamical Astronomy},
  year    = {2017},
  volume  = {128},
  pages   = {383-407},
  adsurl  = {http://adsabs.harvard.edu/abs/2017CeMDA.128..383P},
  doi     = {10.1007/s10569-016-9749-1},
}

@Article{va18,
  author  = {{Voyatzis}, G. and {Antoniadou}, K.~I.},
  title   = {{On quasi-satellite periodic motion in asteroid and planetary dynamics}},
  journal = {Celestial Mechanics and Dynamical Astronomy},
  year    = {2018},
  volume  = {130},
  number  = {9},
  pages   = {59},
  month   = sep,
  doi     = {10.1007/s10569-018-9856-2},
  eid     = {59},
}

@Book{sze,
  title     = {{Theory of orbits. The restricted problem of three bodies}},
  publisher = {Academic Press},
  year      = {1967},
  author    = {{Szebehely}, V.},
  booktitle = {New York: Academic Press, |c1967},
}

@Article{farihi22,
  author        = {{Farihi}, J. and {Hermes}, J.~J. and {Marsh}, T.~R. and {Mustill}, A.~J. and {Wyatt}, M.~C. and {Guidry}, J.~A. and {Wilson}, T.~G. and {Redfield}, S. and {Izquierdo}, P. and {Toloza}, O. and {G{\"a}nsicke}, B.~T. and {Aungwerojwit}, A. and {Kaewmanee}, C. and {Dhillon}, V.~S. and {Swan}, A.},
  title         = {{Relentless and complex transits from a planetesimal debris disc}},
  journal       = {\mnras},
  year          = {2022},
  volume        = {511},
  number        = {2},
  pages         = {1647-1666},
  month         = apr,
  archiveprefix = {arXiv},
  doi           = {10.1093/mnras/stab3475},
}

@Article{vander21,
  author        = {{Vanderbosch}, Zachary P. and {Rappaport}, Saul and {Guidry}, Joseph A. and {Gary}, Bruce L. and {Blouin}, Simon and {Kaye}, Thomas G. and {Weinberger}, Alycia J. and {Melis}, Carl and {Klein}, Beth L. and {Zuckerman}, B. and {Vanderburg}, Andrew and {Hermes}, J.~J. and {Hegedus}, Ryan J. and {Burleigh}, Matthew. R. and {Sefako}, Ramotholo and {Worters}, Hannah L. and {Heintz}, Tyler M.},
  title         = {{Recurring Planetary Debris Transits and Circumstellar Gas around White Dwarf ZTF J0328-1219}},
  journal       = {\apj},
  year          = {2021},
  volume        = {917},
  number        = {1},
  pages         = {41},
  month         = aug,
  archiveprefix = {arXiv},
  doi           = {10.3847/1538-4357/ac0822},
  eid           = {41},
  eprint        = {2106.02659},
  primaryclass  = {astro-ph.EP},
}

@Article{vander20,
  author  = {{Vanderbosch}, Z. and {Hermes}, J.~J. and {Dennihy}, E. and {Dunlap}, B.~H. and {Izquierdo}, P. and {Tremblay}, P. -E. and {Cho}, P.~B. and {G{\"a}nsicke}, B.~T. and {Toloza}, O. and {Bell}, K.~J. and {Montgomery}, M.~H. and {Winget}, D.~E.},
  title   = {{A White Dwarf with Transiting Circumstellar Material Far outside the Roche Limit}},
  journal = {\apj},
  year    = {2020},
  volume  = {897},
  number  = {2},
  pages   = {171},
  month   = jul,
  doi     = {10.3847/1538-4357/ab9649},
  eid     = {171},
}

@Article{vander15,
  author  = {{Vanderburg}, Andrew and {Johnson}, John Asher and {Rappaport}, Saul and {Bieryla}, Allyson and {Irwin}, Jonathan and {Lewis}, John Arban and {Kipping}, David and {Brown}, Warren R. and {Dufour}, Patrick and {Ciardi}, David R. and {Angus}, Ruth and {Schaefer}, Laura and {Latham}, David W. and {Charbonneau}, David and {Beichman}, Charles and {Eastman}, Jason and {McCrady}, Nate and {Wittenmyer}, Robert A. and {Wright}, Jason T.},
  title   = {{A disintegrating minor planet transiting a white dwarf}},
  journal = {\nat},
  year    = {2015},
  volume  = {526},
  number  = {7574},
  pages   = {546-549},
  month   = oct,
  doi     = {10.1038/nature15527},
}

@Article{rapp16,
  author        = {{Rappaport}, S. and {Gary}, B.~L. and {Kaye}, T. and {Vanderburg}, A. and {Croll}, B. and {Benni}, P. and {Foote}, J.},
  title         = {{Drifting asteroid fragments around WD 1145+017}},
  journal       = {\mnras},
  year          = {2016},
  volume        = {458},
  number        = {4},
  pages         = {3904-3917},
  month         = jun,
  archiveprefix = {arXiv},
  doi           = {10.1093/mnras/stw612},
  eprint        = {1602.00740},
  primaryclass  = {astro-ph.EP},
}

@Article{ave16,
  author        = {{Antoniadou}, Kyriaki I. and {Veras}, Dimitri},
  title         = {{Linking long-term planetary N-body simulations with periodic orbits: application to white dwarf pollution}},
  journal       = {\mnras},
  year          = {2016},
  volume        = {463},
  number        = {4},
  pages         = {4108-4120},
  month         = dec,
  archiveprefix = {arXiv},
  doi           = {10.1093/mnras/stw2264},
  primaryclass  = {astro-ph.EP},
}

@Article{ave19,
  author        = {{Antoniadou}, Kyriaki I. and {Veras}, Dimitri},
  title         = {{Driving white dwarf metal pollution through unstable eccentric periodic orbits}},
  journal       = {\aap},
  year          = {2019},
  volume        = {629},
  pages         = {A126},
  month         = sep,
  archiveprefix = {arXiv},
  doi           = {10.1051/0004-6361/201935996},
  eid           = {A126},
  eprint        = {1908.06108},
  primaryclass  = {astro-ph.EP},
}

@PhdThesis{thesis,
  author = {{Antoniadou}, Kyriaki I.},
  title  = {{Dynamics of planetary systems in resonances: From the planar to the spatial}},
  school = {Aristotle University of Thessaloniki, Greece},
  year   = {2014},
  doi    = {10.12681/eadd/34917},
  url    = {https://freader.ekt.gr/eadd/index.php?doc=34917&lang=en},
}

@Article{Nesvorny02,
  author  = {{Nesvorn{\'y}}, D. and {Thomas}, F. and {Ferraz-Mello}, S. and {Morbidelli}, A.},
  title   = {{A Perturbative Treatment of The Co-Orbital Motion}},
  journal = {Celestial Mechanics and Dynamical Astronomy},
  year    = {2002},
  volume  = {82},
  pages   = {323-361},
}

@Article{liwi20,
  author  = {{Liberato}, L. and {Winter}, O.~C.},
  title   = {{The structure of the co-orbital stable regions as a function of the mass ratio}},
  journal = {\mnras},
  year    = {2020},
  volume  = {496},
  number  = {3},
  pages   = {3700-3707},
  month   = aug,
  doi     = {10.1093/mnras/staa1727},
}

@Book{murray,
  title     = {{Solar system dynamics}},
  publisher = {Cambridge University Press},
  year      = {1999},
  author    = {{Murray}, C.~D. and {Dermott}, S.~F.},
  booktitle = {Solar system dynamics by Murray, C.~D., 1999},
}

@Article{rap18,
  author  = {{Rappaport}, S. and {Gary}, B.~L. and {Vanderburg}, A. and {Xu}, S. and {Pooley}, D. and {Mukai}, K.},
  title   = {{WD 1145+017: optical activity during 2016-2017 and limits on the X-ray flux}},
  journal = {\mnras},
  year    = {2018},
  volume  = {474},
  number  = {1},
  pages   = {933-946},
  doi     = {10.1093/mnras/stx2663},
}

@ARTICLE{gurr17,
       author = {{Gurri}, Pol and {Veras}, Dimitri and {G{\"a}nsicke}, Boris T.},
        title = "{Mass and eccentricity constraints on the planetary debris orbiting the white dwarf WD 1145+017}",
      journal = {\mnras},
         year = 2017,
        month = jan,
       volume = {464},
       number = {1},
        pages = {321-328},
          doi = {10.1093/mnras/stw2293},
archivePrefix = {arXiv},
       eprint = {1609.02563},
 primaryClass = {astro-ph.EP},
       adsurl = {https://ui.adsabs.harvard.edu/abs/2017MNRAS.464..321G}
}

@ARTICLE{ver17,
       author = {{Veras}, Dimitri and {Carter}, Philip J. and {Leinhardt}, Zo{\"e} M. and {G{\"a}nsicke}, Boris T.},
        title = "{Explaining the variability of WD 1145+017 with simulations of asteroid tidal disruption}",
      journal = {\mnras},
         year = 2017,
        month = feb,
       volume = {465},
       number = {1},
        pages = {1008-1022},
          doi = {10.1093/mnras/stw2748},
archivePrefix = {arXiv},
       eprint = {1610.06926},
 primaryClass = {astro-ph.EP},
       adsurl = {https://ui.adsabs.harvard.edu/abs/2017MNRAS.465.1008V}
}

@ARTICLE{brou22,
       author = {{Brouwers}, Marc G. and {Bonsor}, Amy and {Malamud}, Uri},
        title = "{A road-map to white dwarf pollution: tidal disruption, eccentric grind-down, and dust accretion}",
      journal = {\mnras},
         year = 2022,
        month = jan,
       volume = {509},
       number = {2},
        pages = {2404-2422},
          doi = {10.1093/mnras/stab3009},
archivePrefix = {arXiv},
       eprint = {2110.07601},
 primaryClass = {astro-ph.EP},
       adsurl = {https://ui.adsabs.harvard.edu/abs/2022MNRAS.509.2404B}
}

@Article{deprit67,
  author  = {{Deprit}, Andr{\'e} and {Jacques}, Henrard and {Palmore}, Julian and {Price}, J.~F.},
  title   = {{The trojan manifold in the system Earth-Moon}},
  journal = {\mnras},
  year    = {1967},
  volume  = {137},
  pages   = {311},
  doi     = {10.1093/mnras/137.3.311},
}

@Article{rein2012,
  author  = {{Rein}, H. and {Liu}, S.~F.},
  title   = {{REBOUND: an open-source multi-purpose N-body code for collisional dynamics}},
  journal = {A\&A},
  year    = {2012},
  volume  = {537},
  pages   = {10pp},
  eid     = {A128},
}

@Article{rein2015,
  author  = {{Rein}, H. and {Spiegel}, D.~S.},
  title   = {{IAS15: a fast, adaptive, high-order integrator for gravitational dynamics, accurate to machine precision over a billion orbits}},
  journal = {MNRAS},
  year    = {2015},
  volume  = {446},
  pages   = {1424-1437},
}

@Article{ave24,
  author  = {{Antoniadou}, Kyriaki I. and {Veras}, Dimitri},
  journal = {\aap},
  title   = {{Disruption of exo-asteroids around white dwarfs and the release of dust particles in debris rings in co-orbital motion}},
  year    = {2024},
  month   = oct,
  pages   = {A249},
  volume  = {690},
  doi     = {10.1051/0004-6361/202451714},
  eid     = {A249},
}

@Article{vta18,
  author  = {{Voyatzis}, G. and {Tsiganis}, K. and {Antoniadou}, K.~I.},
  journal = {Celestial Mechanics and Dynamical Astronomy},
  title   = {{Inclined asymmetric librations in exterior resonances}},
  year    = {2018},
  month   = apr,
  number  = {4},
  pages   = {29},
  volume  = {130},
  doi     = {10.1007/s10569-018-9821-0},
  eid     = {29},
}

@Article{ave26a,
  author  = {{Antoniadou}, Kyriaki I. and {Veras}, Dimitri},
  journal = {\aap},
  year    = {2026a},
  note    = {Accepted},
  doi     = {10.1051/0004-6361/202661508},
}

@Article{mich79c,
  author  = {{Michalodimitrakis}, M.},
  journal = {Astronomy and Astrophysics},
  title   = {{A contribution to the three-dimensional Copenhagen problem}},
  year    = {1979},
  pages   = {6-14},
  volume  = {76},
}

@Article{Deprit1969,
  author  = {{Deprit}, A. and {Henrard}, J.},
  journal = {\aap},
  title   = {{On the Periodical Orbits Emanating from L4 in the Resonance Case 1/4}},
  year    = {1969},
  month   = sep,
  pages   = {88},
  volume  = {3},
}

@Article{Henrard1970,
  author  = {{Henrard}, J.},
  journal = {\aap},
  title   = {{Concerning the Genealogy of Long Period Families at L4}},
  year    = {1970},
  month   = mar,
  pages   = {45},
  volume  = {5},
}

@Article{Henrard1970a,
  author   = {{Henrard}, Jacques},
  journal  = {Celestial Mechanics},
  title    = {{Periodic Orbits Emanating from a Resonant Equilibrium}},
  year     = {1970},
  month    = sep,
  number   = {3-4},
  pages    = {437-466},
  volume   = {1},
  doi      = {10.1007/BF01231143},
}

@Article{Bien1978,
  author  = {{Bien}, R.},
  journal = {\aap},
  title   = {{Long-Period Effects in the Motion of Trojan Asteroids and of Fictitious Objects at the 1/1 Resonance. Fictitious Objects}},
  year    = {1978},
  month   = aug,
  number  = {1-2},
  pages   = {295-301},
  volume  = {68},
}

@Article{Lohinger1993,
  author  = {{Lohinger}, E. and {Dvorak}, R.},
  journal = {\aap},
  title   = {{Stability regions around L4 in the elliptic restricted problem}},
  year    = {1993},
  month   = dec,
  number  = {2},
  pages   = {683-687},
  volume  = {280},
}

@Article{Erdi2007,
  author  = {{{\'E}rdi}, B. and {Nagy}, I. and {S{\'a}ndor}, Zs. and {S{\"u}li}, {\'A}. and {Fr{\"o}hlich}, G.},
  journal = {\mnras},
  title   = {{Secondary resonances of co-orbital motions}},
  year    = {2007},
  month   = oct,
  number  = {1},
  pages   = {33-40},
  volume  = {381},
  doi     = {10.1111/j.1365-2966.2007.12228.x},
}

@Article{arnold,
  author  = {{Arnol'd}, V.~I.},
  journal = {Russian Mathematical Surveys},
  title   = {{Proof of a Theorem of A. N. KOLMOGOROV on the Invariance of Quasi-Periodic Motions Under Small Perturbations of the Hamiltonian}},
  year    = {1963},
  pages   = {9-36},
  volume  = {18},
}

@InProceedings{berry,
  author    = {{Berry}, M.~V.},
  booktitle = {American Institute of Physics Conference Series},
  title     = {{Regular and irregular motion}},
  year      = {1978},
  pages     = {16-120},
  series    = {American Institute of Physics Conference Series},
  volume    = {46},
}

\onecolumn
\begin{appendix}
\section{Bifurcation points generating orbits in P5}\label{app1}
\begin{table*}[!h]
\centering
\caption{Bifurcation points that justify the origin and continuation of 1/1 resonant periodic orbits in the 2D-ERTBP and the 3D-ERTBP presented within the Ceres-Neptune domain.}
\begin{tabular}[b]{lccccccccc}
\toprule
& $e_{\rm a}$ &$e_{\rm d}$&$\theta$  &$\varpi_{\rm a} (^{\circ})$     &$M_{\rm a} (^{\circ})$ &$\varpi_{\rm d} (^{\circ})$  &$M_{\rm d} (^{\circ})$& \begin{tabular}{@{}c@{}}Exists in \\ Problem/ \\ Plot\end{tabular} & \begin{tabular}{@{}c@{}}Generates \\orbits in \\ Problem/Plot\end{tabular}\\
\cmidrule{1-10}
$B^1_{T}$    &0      &0.835   &0     &0       &0   &180   &180 & \begin{tabular}{@{}c@{}}2D-CRTBP \end{tabular} & \begin{tabular}{@{}c@{}}2D-ERTBP \\ Fig. \ref{2DE}a\end{tabular}\\
\cmidrule{1-10}
$B^2_{T}$    &0      &0.977   &180     &0       &0   &0   &180 & \begin{tabular}{@{}c@{}}2D-CRTBP \end{tabular} & \begin{tabular}{@{}c@{}}2D-ERTBP \\ Fig. \ref{2DE}a\end{tabular}\\
\cmidrule{1-10}
$B^3_{T}$    &0      &0.869   &146.74     &180       &180   &133.43   &79.83 & \begin{tabular}{@{}c@{}}2D-CRTBP \end{tabular} & \begin{tabular}{@{}c@{}}2D-ERTBP \\ Fig. \ref{2DE}a\end{tabular}\\
\cmidrule{1-10}
$B^4_{T}$    &0      &0.964   &167.19     &180       &180   &314.66   &212.53 & \begin{tabular}{@{}c@{}}2D-CRTBP \end{tabular} & \begin{tabular}{@{}c@{}}2D-ERTBP \\ Fig. \ref{2DE}b,c\end{tabular}\\
\cmidrule{1-10}
$B^5_{T}$    &0      &\begin{tabular}{@{}c@{}}0.815 \\ (when $i_{\rm d}=22.5^{\circ}$)\end{tabular}   &0      &\begin{tabular}{@{}c@{}}0 \\ (with $\Omega_{\rm a}=0^{\circ}$\\ and $\omega_{\rm a}=0^{\circ}$)\end{tabular} &0       &\begin{tabular}{@{}c@{}}180 \\ (with $\Omega_{\rm d}=270^{\circ}$\\ and $\omega_{\rm d}=270^{\circ}$)\end{tabular}
   &180 & \begin{tabular}{@{}c@{}}3D-CRTBP \end{tabular} & \begin{tabular}{@{}c@{}}3D-ERTBP \\ Fig. \ref{3DE}\end{tabular}  \\
   \cmidrule{1-10}
$B^1_{es}$    &0.087      &0.798   &0     &180       &180   &0   &0 &  \begin{tabular}{@{}c@{}}2D-ERTBP \\ Fig. \ref{2DE}a \end{tabular} & \begin{tabular}{@{}c@{}}3D-ERTBP \\ Fig. \ref{3DE}\end{tabular}\\
\bottomrule
\end{tabular}\label{tab1}
\tablefoot{As the order of $\mu$ varies from Ceres to Neptune, no significant change was reflected on $e_{\rm d}$ (order of $10^{-4}$ for all bifurcation points). The plots regarding the 2D-CRTBP and the 3D-CRTBP are included in the respective Table by \citet{ave26a} in study P4.}
\end{table*}

\section{Variation of orbital elements when $\mu\in[10^{-10},10^{-5}]$ in the 2D-ERTBP}\label{app1b}
\begin{figure*}[!h]\centering
\setlength{\arraycolsep}{0.5pt}
$\begin{array}{cccc}
  \includegraphics[width=0.25\textwidth]{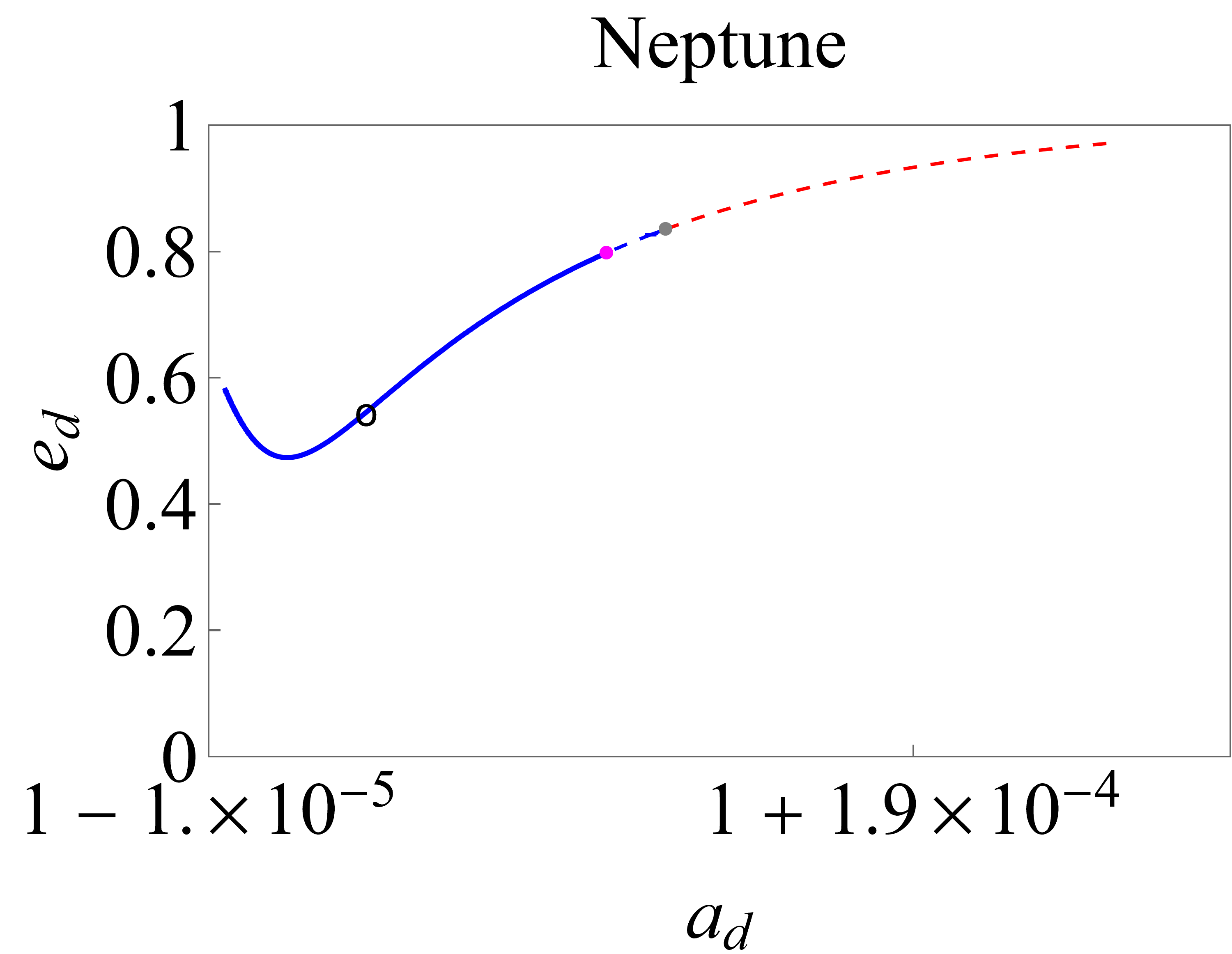}&\includegraphics[width=0.25\textwidth]{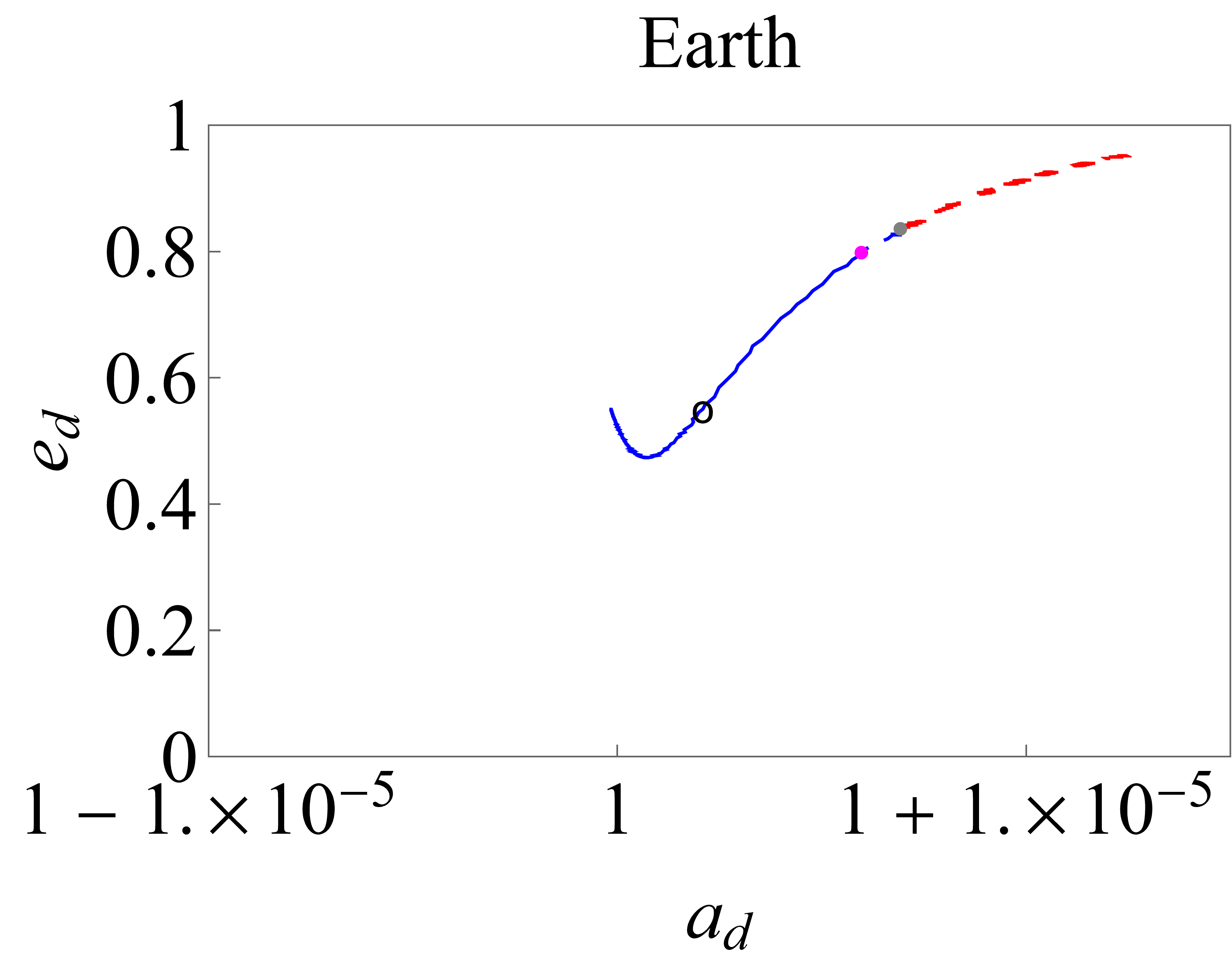}&\includegraphics[width=0.25\textwidth]{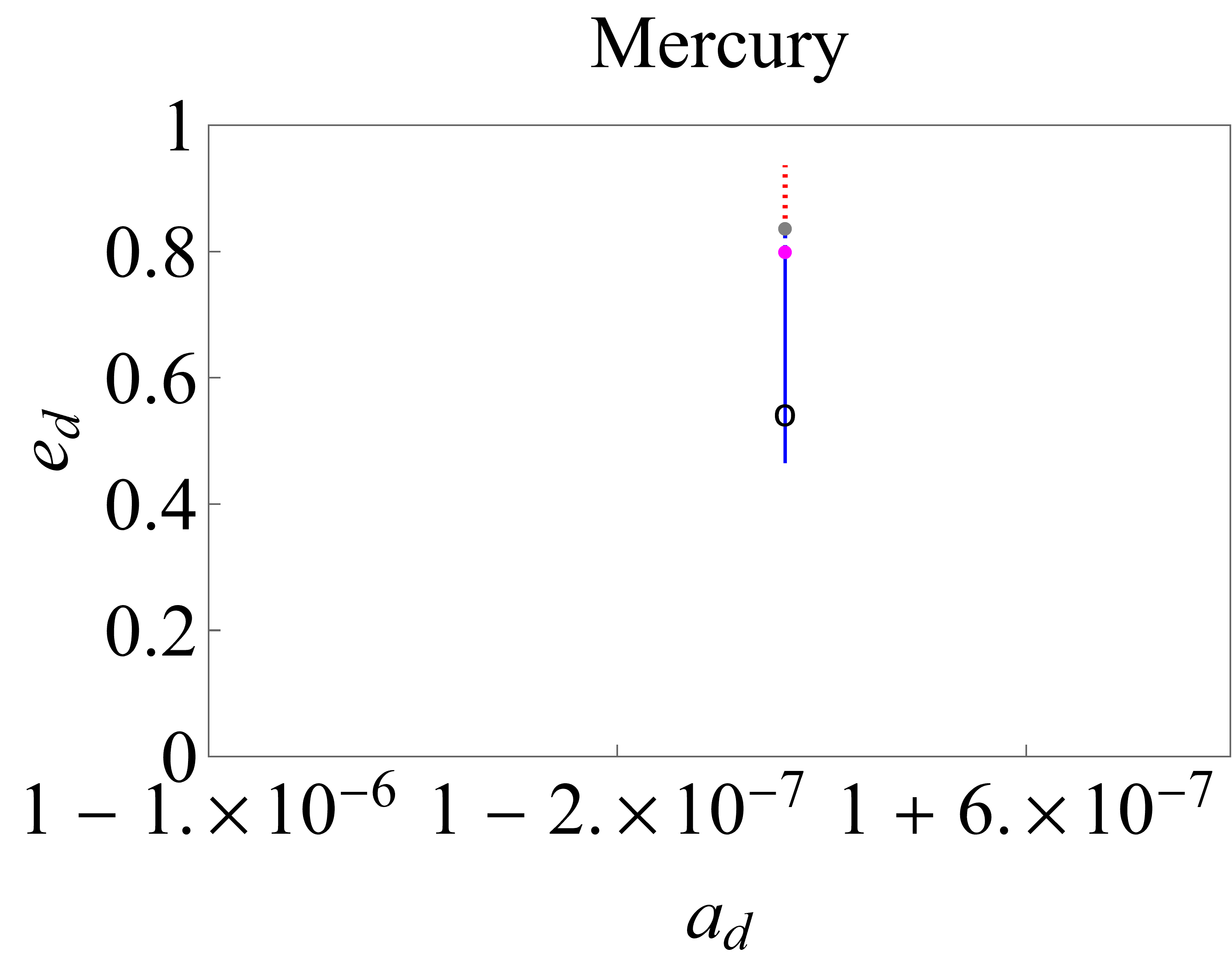}&\includegraphics[width=0.25\textwidth]{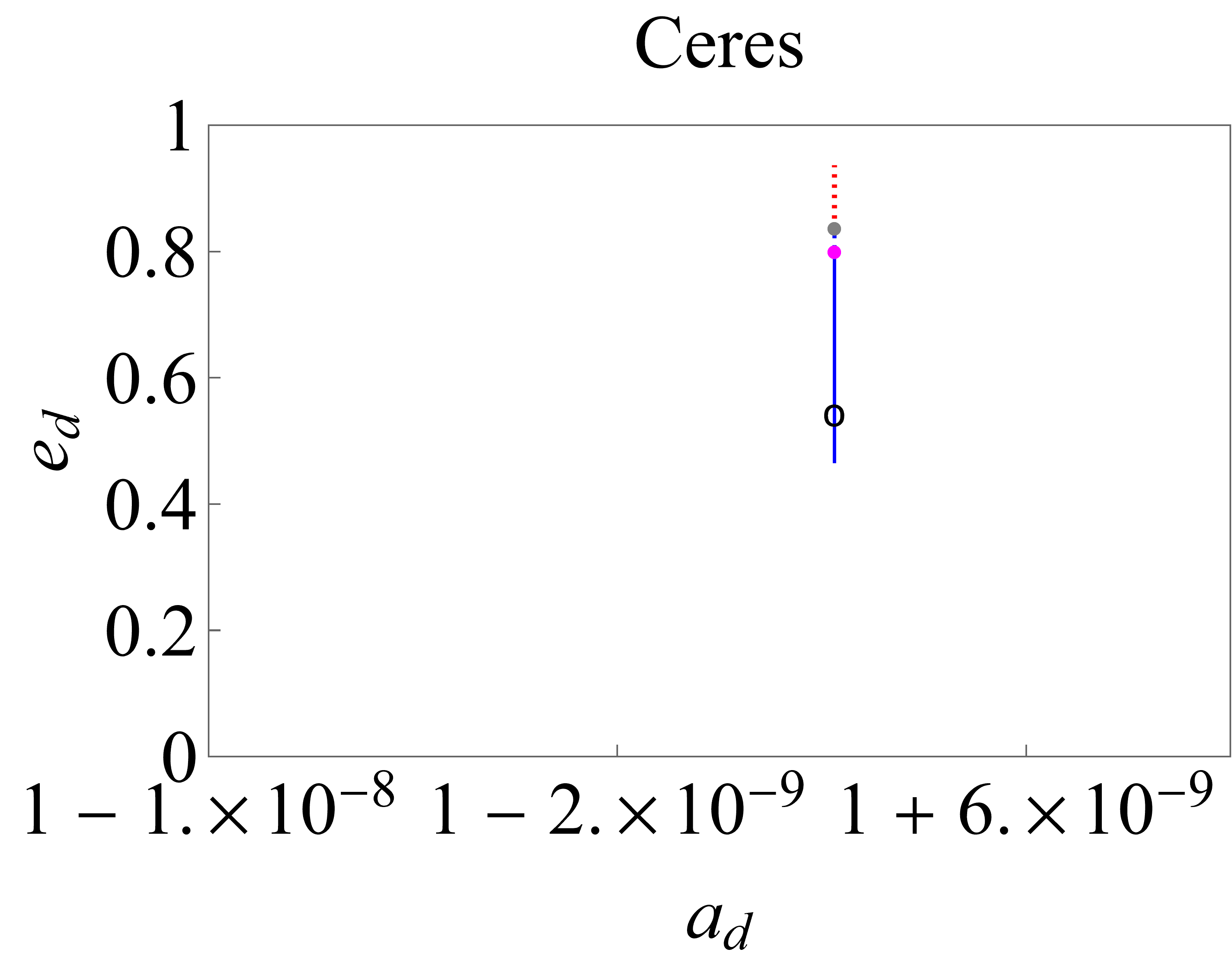}\\
\end{array}$
\caption{$E_{1a}$ family is showcased on the $(a_{\rm d}, e_{\rm d})$ plane. The minute changes encountered at $a_{\rm d}$ are of the order of $\mu$. The black circle corresponds to the periodic orbit chosen for the DS maps and the simulations. Presentation as in Fig. \ref{2DE}.}
        \label{E1a_cla}
\end{figure*}

\twocolumn
\section{Families of spatial symmetric periodic orbits in the 3D-ERTBP}\label{app2}
\begin{figure}[!h]\centering
$\begin{array}{c}
  \includegraphics[width=0.35\textwidth]{3DE.png}\\
\end{array}$
\caption{\textit{Panel (a)}: Families of spatial symmetric periodic orbits in the $(e_{\rm a},e_{\rm d},i_{\rm d})$ space in the 3D-ERTBP with $a_{\rm a}=1$ and $i_{\rm a}=0^{\circ}$. The 3D family $F_a$ of the 3D-ERTBP consists of stable periodic orbits, starts at $B_{es}$ of the planar $E_{1a}$ family of the 2D-ERTBP (see also Fig. \ref{2DE}a) and terminates at $B^5_T$ found at $F^I$ of the 3D-CRTBP (see P4). The other branch of unstable periodic orbits in the 3D-ERTBP, called $F_p$ reaches inclination values close to $60^{\circ}$ as $e_{\rm d}\rightarrow 1$. Label C stands for Ceres and N for Neptune. \textit{Panel (b)}: Magnification of panel (a) in the vicinity of the bifurcation points.}
        \label{3DE}
\end{figure}

\begin{figure}[!h]\centering
$\begin{array}{c}
  \includegraphics[width=0.32\textwidth]{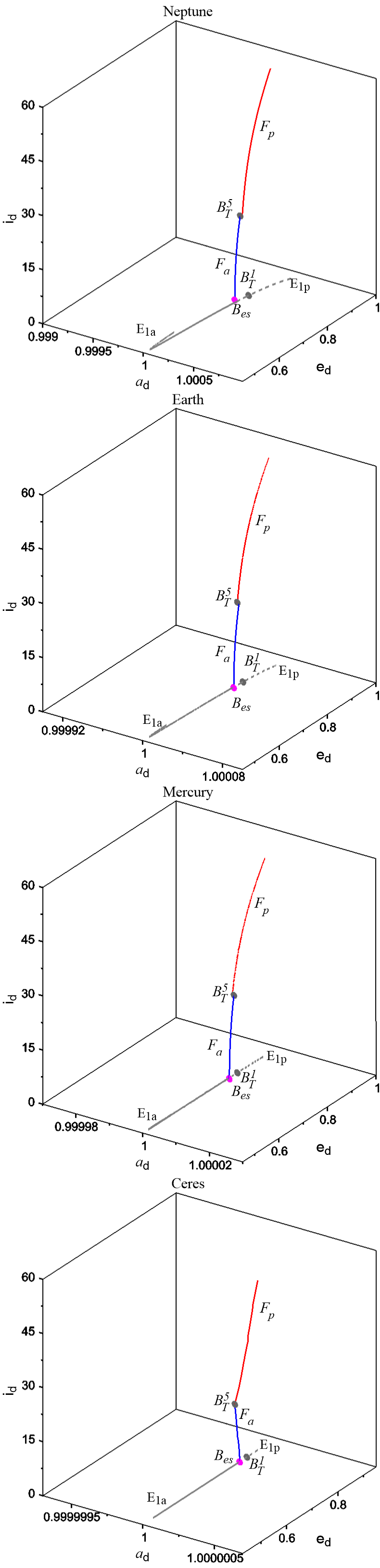}\\\vspace{-0.6cm}
\end{array}$
\caption{Families $F_p$ and $F_a$  in the $(a_{\rm d}, e_{\rm d}, i_{\rm d})$ space. As in the 2D-ERTBP, the minute changes encountered at $a_{\rm d}$ are of the order of $\mu$.}
        \label{3deexp}
\end{figure}
\clearpage

\section{Evolution of angles}\label{app3}
\begin{figure}[!h]\centering
$\begin{array}{c}
  \includegraphics[width=1.\columnwidth]{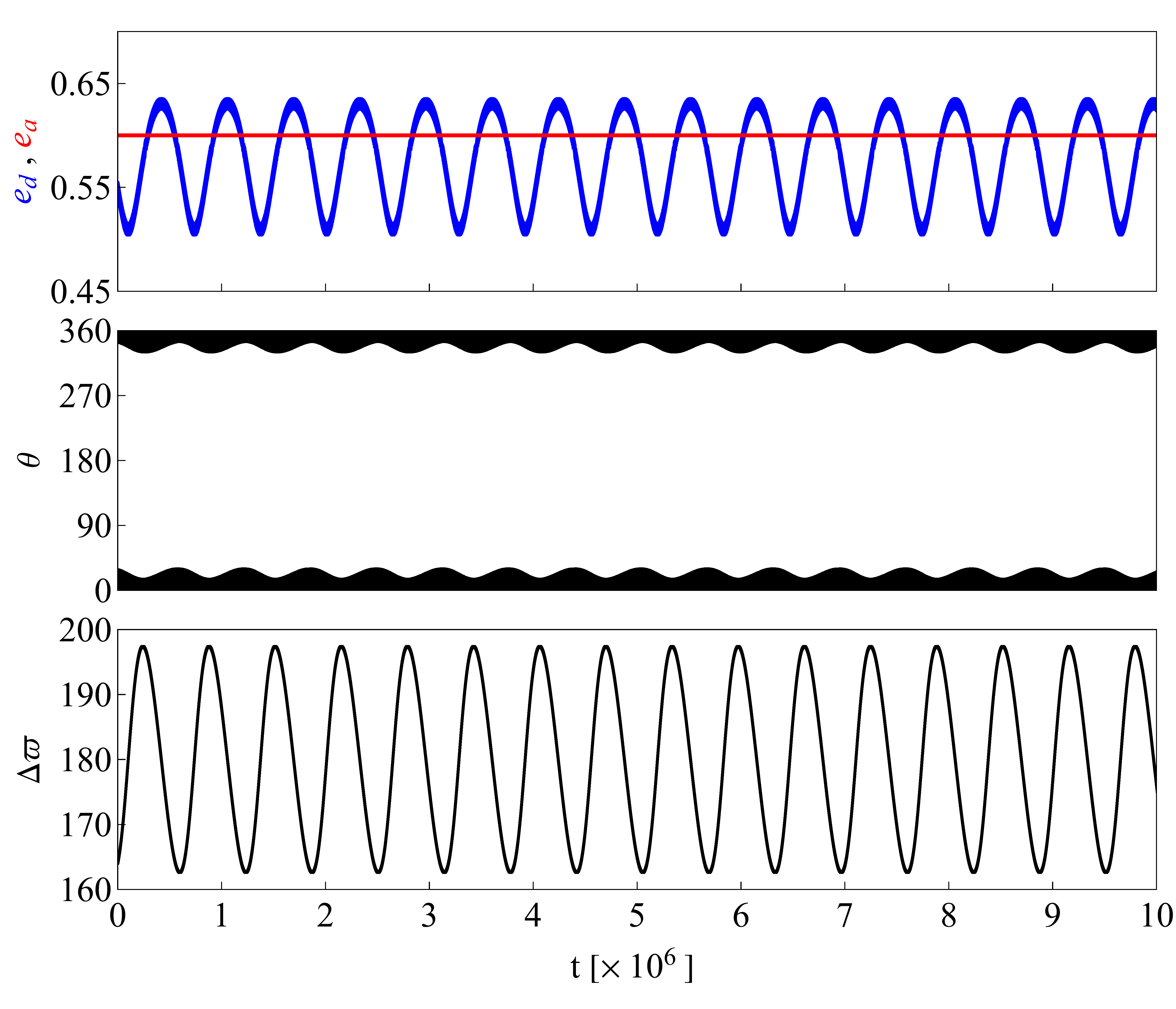}\\
\end{array}$
\caption{Typical evolution about stable resonant planar symmetric periodic orbits found in Fig. \ref{2DE_TB}, where $\theta$ librates about $0^{\circ}$ and the apsidal difference librates about $180^{\circ}$. }
        \label{Bottom_A}
\end{figure}

\begin{figure}[!h]\centering
$\begin{array}{c}
  \includegraphics[width=1.\columnwidth]{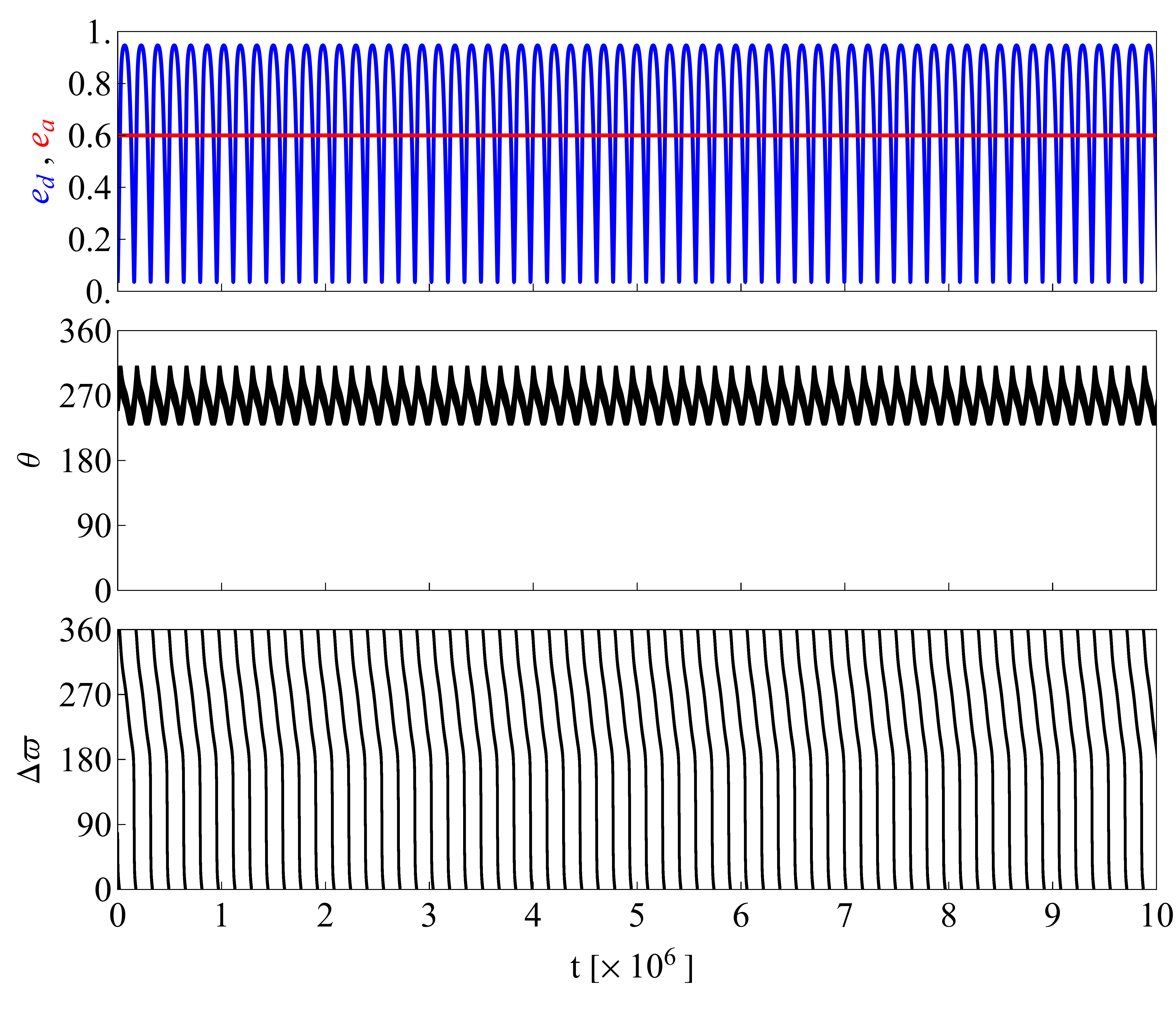}\\
\end{array}$
\caption{Typical evolution with initial conditions extracted from low-eccentricity regimes of Fig. \ref{2DE_TB}, where $e_{\rm d}$ exhibits high amplitude oscillations and the phases are not protected ($\Delta\varpi$ circulates). The libration of $\theta$ accounts for the long-term stability within these regions. }
        \label{Bottom_B}
\end{figure}

\begin{figure}[!h]\centering
$\begin{array}{c}
  \includegraphics[width=1.\columnwidth]{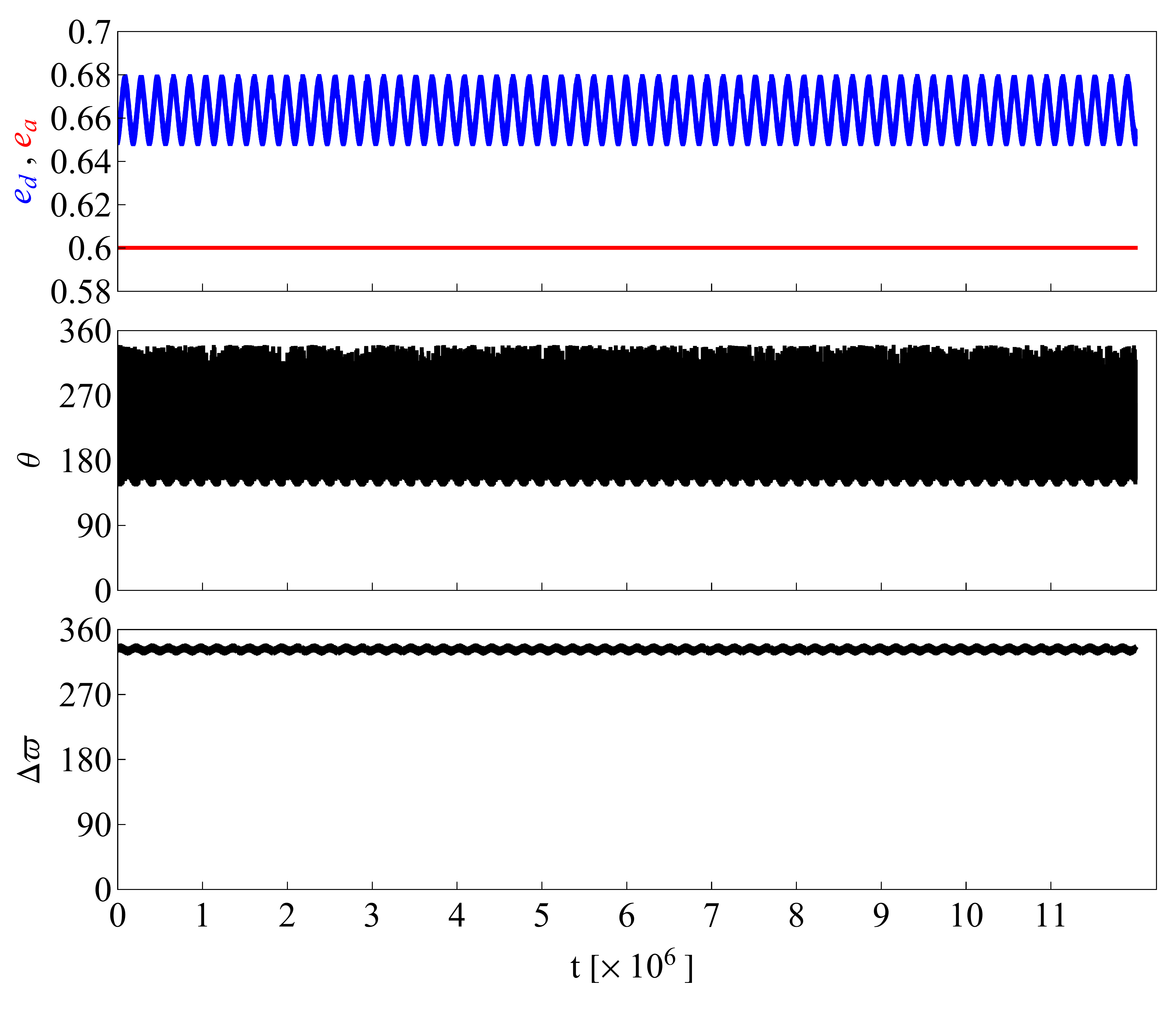}\\
\end{array}$
\caption{Typical evolution around planar stable resonant asymmetric periodic orbits shown in Fig. \ref{earth} with initial conditions extracted from  Fig. \ref{2DE_TB}. All angles librate (MMR) around angles different than $0^{\circ}$ or $180^{\circ}$.}
        \label{Bottom_C}
\end{figure}

\begin{figure}[!h]\centering
$\begin{array}{c}
  \includegraphics[width=1.\columnwidth]{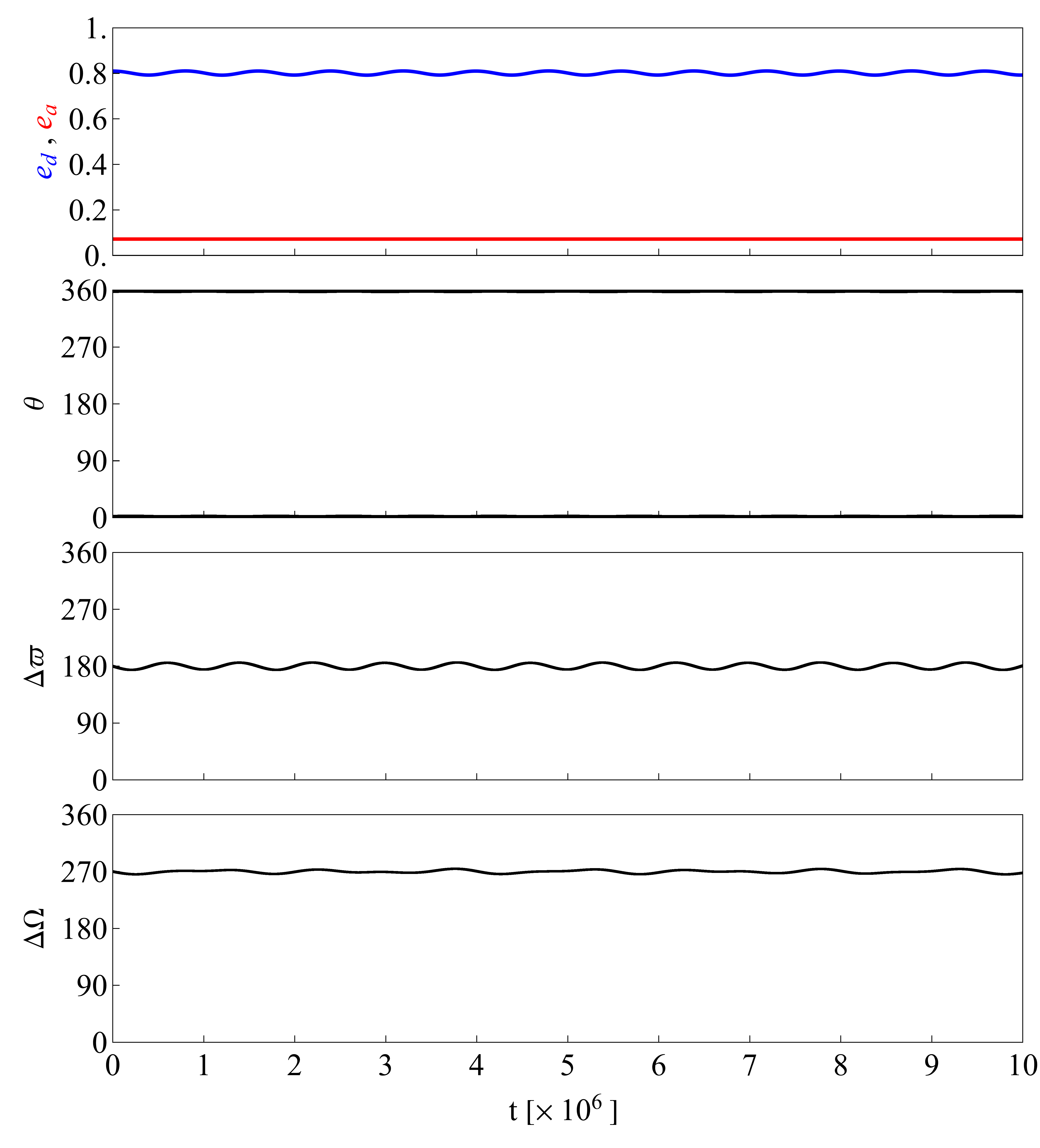}\\
\end{array}$
\caption{Typical orbital evolution, where all angles librate, around spatial stable resonant symmetric periodic orbits shown in Fig. \ref{3DE} with initial conditions extracted from  Fig. \ref{3DE_Di10}.}
        \label{Fa_A}
\end{figure}

\begin{figure}[!h]\centering
$\begin{array}{c}
  \includegraphics[width=1.\columnwidth]{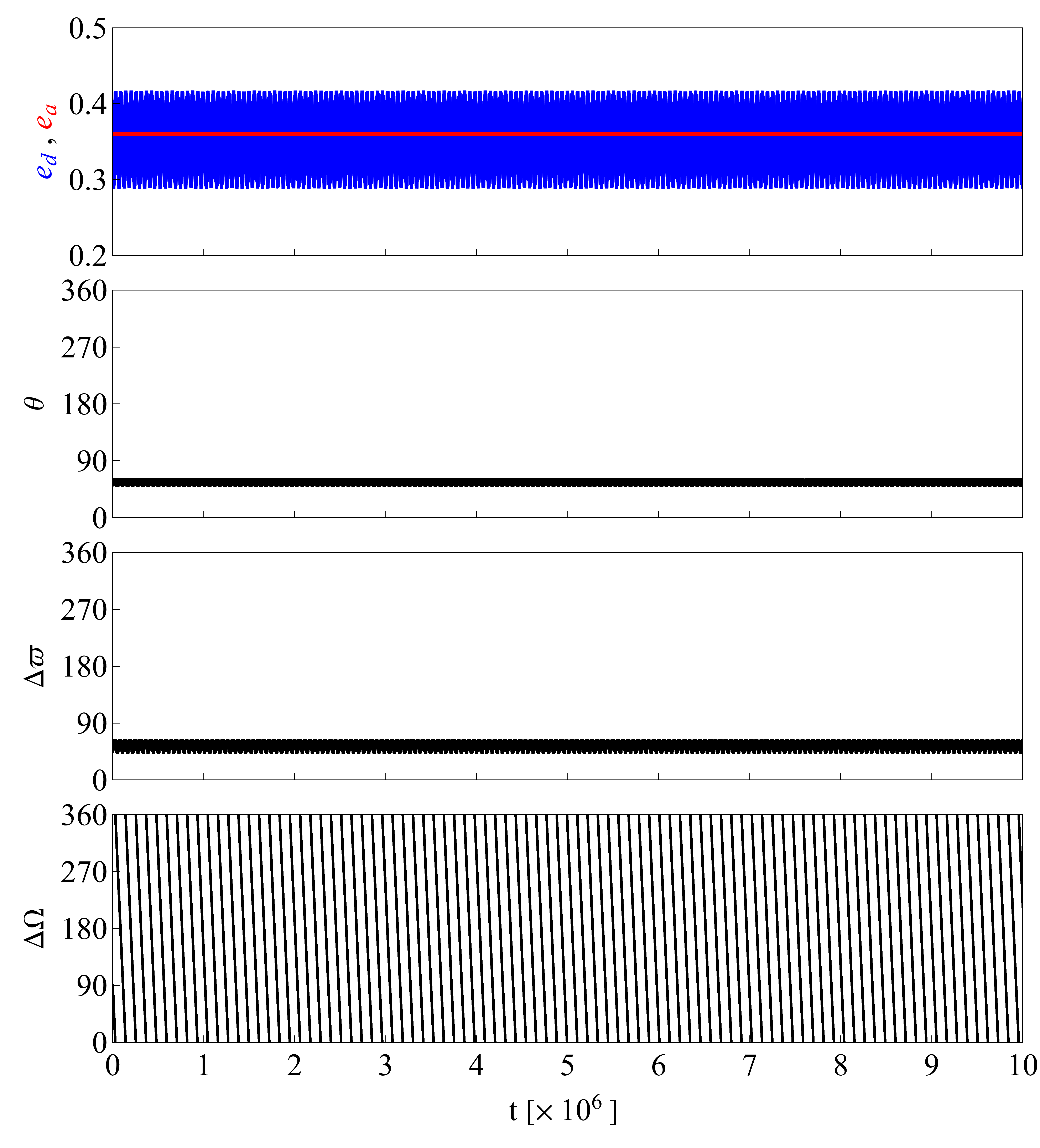}\\
\end{array}$
\caption{Typical orbital evolution of a highly inclined dust particle found around planar asymmetric vertically stable resonant periodic orbits shown in Fig. \ref{aL4} with initial conditions extracted from  Fig. \ref{3DE_Di45}. An MMR is solely exhibited; no i-resonance.}
        \label{Fp_C}
\end{figure}

\begin{figure}[!h]\centering
$\begin{array}{c}
  \includegraphics[width=1.\columnwidth]{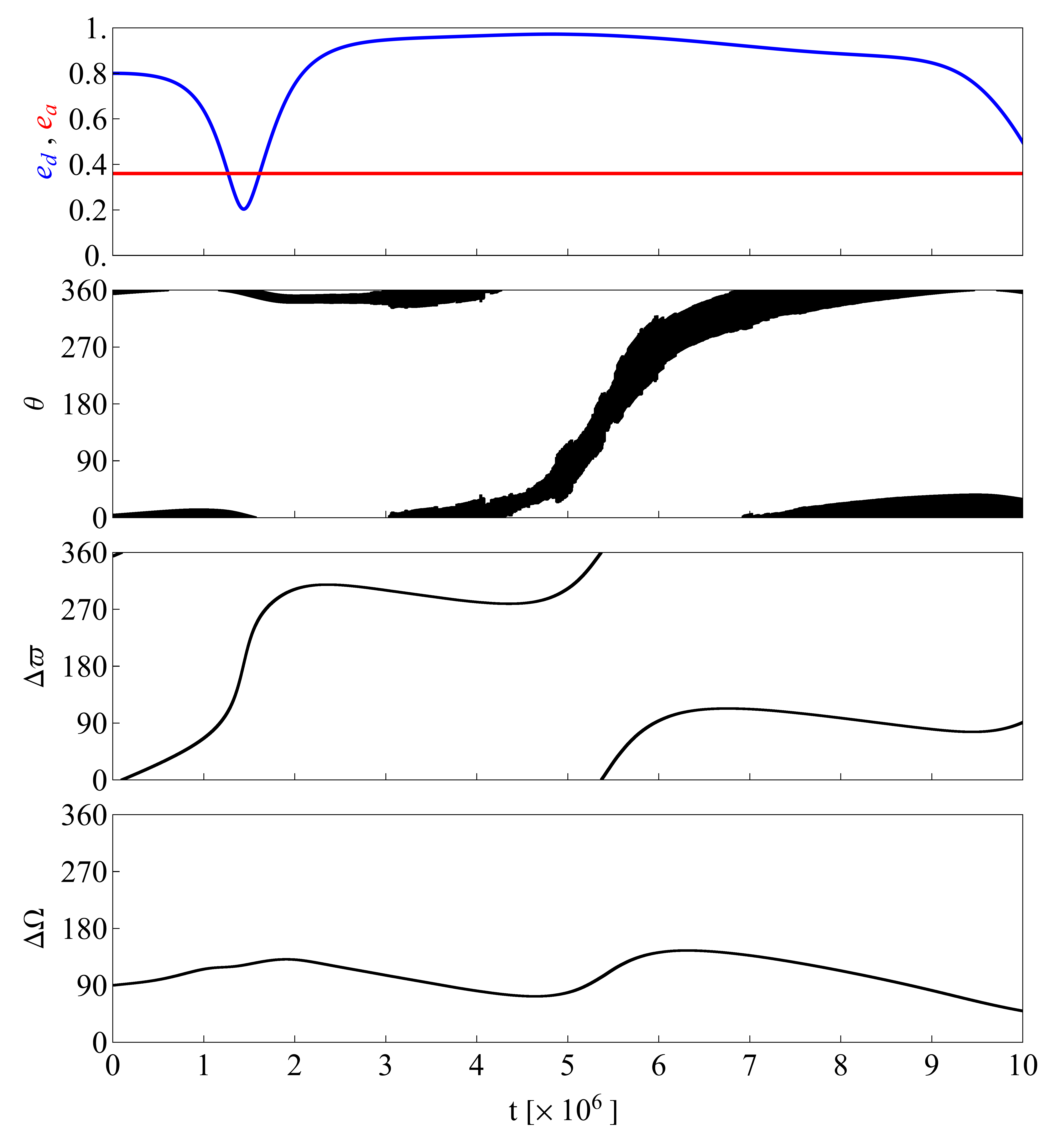}\\
\end{array}$
\caption{Typical orbital evolution around spatial unstable resonant periodic orbits shown in Fig. \ref{3DE} with initial conditions extracted from  Fig. \ref{3DE_Di45}. The nodal difference oscillation safeguards the highly inclined dust particle within these vicinities.}
        \label{Fp_E}
\end{figure}

In Figs. \ref{Bottom_A}-\ref{Bottom_C}, we show typical scenarios safeguarding dust particles around WDs under the gravitational attraction of a secondary moving on an elliptic orbit (2D-ERTBP). The apsidal difference is defined as $\Delta\varpi=\varpi_{\rm a}-\varpi_{\rm d}$ entailing  coplanarity ($i_{\rm a}=i_{\rm d}=\Omega_{\rm a}=\Omega_{\rm d}=0^{\circ}$). Then, in Figs. \ref{Fa_A}-\ref{Fp_E}, we demonstrate the typical scenarios safeguarding inclined dust particles around WDs under the gravitational attraction of a secondary moving on an elliptic orbit (3D-ERTBP).

\end{appendix}

\end{document}